\documentclass[aps,prx,superscriptaddress,twocolumn,preprintnumbers,amsmath,amssymb,aps,longbibliography,showkeys]{revtex4-2} %preprint

\usepackage{graphicx} % Required for inserting images
\usepackage{bm}
\usepackage{xcolor}
\usepackage{tabularx}

\usepackage{hyperref}
\hypersetup{
    colorlinks=true,
    linkcolor=red,
    filecolor=magenta,      
    urlcolor=cyan,
    citecolor = magenta,
    pdftitle={Inference of time delay in stochastic systems},
    }

\begin{document}
\title{Inference of a time delay in stochastic systems}

\author{Robin A. Kopp}
\email{r.kopp@tu-berlin.de}
\affiliation{Institut für Physik und Astronomie, Technische Universität Berlin, Hardenbergstraße 36, D-10623 Berlin, Germany}
\author{Sabine H. L. Klapp}
\email{sabine.klapp@tu-berlin.de}
\affiliation{Institut für Physik und Astronomie, Technische Universität Berlin, Hardenbergstraße 36, D-10623 Berlin, Germany}
\author{Deepak Gupta}
\email{phydeepak.gupta@gmail.com}
\affiliation{Institut für Physik und Astronomie, Technische Universität Berlin, Hardenbergstraße 36, D-10623 Berlin, Germany}

\date{\today}

\begin{abstract}
Time delay is ubiquitous in many experimental and real-world situations. However, from measurable data it is often unclear whether time delay plays a significant role in the dynamics,
and if it does, how long the time lag really is. This would be invaluable knowledge when analyzing and modeling such systems. Hitherto, no universal method is available by which the time delay can be inferred. 
To address this problem, we propose and demonstrate two different methods to infer time delay in overdamped Langevin systems with delayed feedback. In the first part, we focus on the power spectral density based on the positional data and use a characteristic signature of the time delay to infer the delay time. In limiting cases, we establish a direct relation of the observations made for nonlinear time-delayed feedback forces to analytical results obtained for the linear system. 
In other situations 
despite the absence of this direct relation, the characteristic signature remains and can be exploited by a semiautomatic method to infer the delay time. 
Furthermore, it 
may not always desirable or possible to observe a system for a long time to infer dependencies and parameters. 
Thus, in the second part, we propose 
a probing method combined with a neural network to infer the delay time, which requires only short observation time series.
These proposed methods for inferring time delays in stochastic systems 
can
be valuable tools for gaining deeper insight into the role of delay across a wide range of applications -- from the behavior of individual colloidal particles under feedback control to emergent collective phenomena such as flocking and swarming.

\end{abstract}

%\keywords{}

\maketitle

\section{Introduction}
Time delay can have a substantial impact on a wide range of experimental and real-world scenarios~\cite{schollControlSelforganizingNonlinear2016}.
The presence of time delay makes the dynamics of a system dependent on 
the system's history itself.
In some cases, it is introduced deliberately, whereas in other situations 
its presence is unavoidable. 
A prominent example for deliberate use of delay is
Pyragas time-delayed feedback for the stabilization of otherwise unstable orbits \cite{pyragasContinuousControlChaos1992,pyragas_control_1995}, later applied to stabilization of satellite motion~\cite{tsuiControlHigherDimensional2000}. 
In experiments involving feedback control, researchers have explicitly considered %the
time delay 
in design and modeling
~\cite{gavrilovFeedbackTrapsVirtual2017,kumarNanoscaleVirtualPotentials2018,bell-daviesDynamicsColloidalParticle2023}. 
In colloidal systems exhibiting stochastic dynamics, time-delayed feedback control is employed to modify the transport properties
\cite{lichtnerFeedbackcontrolledTransportInteracting2010,lopezRealizationFeedbackControlled2008}. 
Moreover, delay has been shown to have tremendous effect on 
thermodynamic observables
~\cite{loos_heat_2019,koppHeatProductionStochastic2024,munakataLinearStochasticSystem2009,munakataEntropyProductionFluctuation2014a,debiossacThermodynamicsContinuousNonMarkovian2020}, as demonstrated in recent experiments~\cite{bell-daviesDynamicsColloidalParticle2023,debiossacNonMarkovianFeedbackControl2022}.

Another area where delay is important and, at the same time difficult to control, are systems with finite-time 
information
propagation.
Typical examples 
include communication lags 
while steering far 
objects 
such as satellites, the reaction time humans experience when reacting to an external stimulus,
or the time lag in experiments between processing image data and applying the desired steering or control impulse~\cite{franzlFullySteerableSymmetric2021}. 
Moreover, time delay may have to be taken into consideration  
in complex systems, 
such as in swarms \cite{hindesHybridDynamicsDelaycoupled2016}, in financial markets \cite{stoicaStochasticDelayFinancial2004,callenAccountingQualityStock2013}, in robotic systems \cite{mijalkovEngineeringSensorialDelay2016}, and in traffic flow \cite{sipahiStabilityTrafficFlow2008,yuDensityWavesTraffic2010}.
Indeed,
it has recently been shown in theoretical studies as well as simulations that the introduction of time delay and/or time-delayed feedback can 
induce intriguing 
emergent collective behavior.
Examples 
range from 
polar active matter with 
delayed interactions~\cite{holubecFiniteSizeScalingEdge2021,pakpourDelayinducedPhaseTransitions2024}, and Gaussian time-delayed feedback applied to Brownian particles~\cite{koppSpontaneousVelocityAlignment2023,taramaTravelingBandFormation2019,taramaEngineeringLivingWorms2025a},
to ecological systems~\cite{piganiDelayEffectsStability2022},
and oscillator networks \cite{earlSynchronizationOscillatorNetworks2003,dahmsClusterGroupSynchronization2012}. 

Stochastic systems with time delay belong to the class of non-Markovian systems since the dynamics depends on the systems state at an earlier time, i.e., it is history dependent. Fundamentally, the time delay may be rationalized as a consequence of hidden degrees of freedom. From a Markovian embedding perspective \cite{siegleMarkovianEmbeddingNonMarkovian2010}, for a system with discrete time delay, an infinite number of degrees of freedom have been projected out \cite{loosFokkerPlanckEquations2019,doerriesCorrelationFunctionsNonMarkovian2021}. Further, in the context of generalized Langevin equations, discrete time delay systems can be understood as systems with a $\delta$-shaped memory kernel. Notice that this memory kernel, however, is not related to the environmental noise correlation \cite{loosFokkerPlanckEquations2019}.

However, in many experimental and real-world situations, it is not a priori 
clear whether time delay plays a significant role~\cite{franzlActiveParticleFeedback2020,franzlFullySteerableSymmetric2021} or if it is present at all. 
Therefore, it is of fundamental importance to investigate the presence of time delay 
based on observable quantities
to accurately design and understand physical systems.

In this study, we focus on systems with a specific delay time rather than multiple delays or distributed delay \cite{loosMediumEntropyReduction2021}.
Of course, assuming a single discrete time delay in a real system is a rather strong assumption. 
Broadly speaking, systems with several time delays, distributed delay or a more general memory kernel 
have been studied.
However, there are 
numerous 
systems where the discrete time delay assumption is well justified. 
In laser systems \cite{ottoModelingQuantumDot2010,ottoCOMPLEXDYNAMICSSEMICONDUCTOR2012a} and quantum-optical systems \cite{dornerLaserdrivenAtomsHalfcavities2002,carmeleSinglePhotonDelayed2013}, time delay originates from light traveling through the optical setup at finite speed.
Further examples include complex networks with finite signal propagation speeds (see Refs. \cite{schollControlSelforganizingNonlinear2016,atayComplexTimeDelaySystems2010} and references therein on the topics of neural and laser networks). 
In other cases, one can expect a `typical' time delay as part of a peaked distribution, e.g., when considering finite processing and reaction times. Therefore, gaining information about the presence of time delay as well as the `typical' or average time delay is of high value. 

In recent years, ideas of parameter inference for dynamical systems and equation discovery (for ordinary as well as partial differential equations) from (noisy) data have received considerable interest. Methods range from Bayesian inference  
\cite{vontoussaintBayesianInferencePhysics2011} and regression-based machine learning \cite{bruntonDiscoveringGoverningEquations2016}  
to the application of neural network-based techniques \cite{chenPhysicsinformedLearningGoverning2021}. 
For some types of system, methods for identifying governing equations \cite{kopeczi-boczDataDrivenDelayIdentification2024,leylazIdentificationNonlinearDynamical2022} and inferring parameters for deterministic and linear stochastic delay differential equations (DDE and SDDE) \cite{zhaoInferenceDelayDifferential2024, kuchlerStatisticalInferenceDiscretetime2013}, specifically time delay in the underlying dynamics  from time series data \cite{vossReconstructionNonlinearTime1997,horbeltParameterEstimationNonlinear2002,bunnerToolRecoverScalar1996} exist. However, those methods are typically designed for application to deterministic systems or noisy measurement data \cite{horbeltParameterEstimationNonlinear2002}. For fully stochastic (non)linear systems, on the other hand, to the best of our knowledge, no universal approach for the inference of time delay
exists.
Motivated by this gap,
we 
here
propose and demonstrate, for the first time, two approaches to infer 
a 
time delay 
for Brownian systems. 

The first approach is based on the power spectral density (PSD) of stochastic time series, which is often used to analyze experimental data \cite{muenkerAccessingActivityViscoelastic2024,berg-sorensen_power_2004,berg-sorensen_power_2006} including systems with anomalous diffusion \cite{sposiniRobustCriterionAnomalous2022}.
Here, we
identify a characteristic signature of time delay, 
namely 
oscillations  
with a delay-specific period, 
and demonstrate in a proof of concept how this signature can be exploited to infer the delay time. 
To do this, we start from
Brownian systems subject to linear time-delayed feedback as well as specific delayed nonlinear forces including respective limiting cases 
and then consider
highly nonlinear time-delayed feedback.
We 
thereby
demonstrate 
that
the characteristic signature 
carries over from systems subject to linear feedback forces to nonlinear systems
and, thus, that 
the PSD-based inference approach is agnostic, i.e., it does not rely on knowledge 
of 
the functional form of the underlying stochastic equations of motion. 

The PSD approach, however, has a drawback. A large number (typically $N\gtrsim 100$) of relatively long [several Brownian times $\tau_B$~(defined in Sec.~\ref{sec:model_basic})] trajectories is 
required to infer 
the time delay with sufficient precision. 
In real-world applications,
having to record a large number of long trajectories may (depending on the scenario) not be desirable or not even possible.

Given the potential difficulties, we propose a powerful alternative approach
employing a neural network.
To establish the concept,
we utilize the fact
that systems that involve time delay, rather than instantaneously responding to an external stimulus, react with some delay. Therefore, by applying a suitable perturbation and observing the dynamical response, 
we can obtain information about the underlying delay time of the system. 
In order to classify the system's response, based on trajectory time series data, we make use of a convolutional neural network \cite{brunton_data-driven_2022,fawaz_deep_2019}.
Such a network  can be trained to output the most likely time delay for one or even 
several different functional forms of the underlying delayed and instantaneous forces.
Compared
to the PSD approach, 
the main advantage of this approach is that the observation window (i.e., the trajectory length) only has to be of the order of the time delay itself, and relatively few trajectories (typically $N\sim25$) are sufficient for precise inference. In combination with 
state-of-the-art
computer systems, this brings the inference of time delay in real time within the realm of possibilities.

The remainder of this paper is structured as follows.
In Sec.~\ref{sec:model_basic}, we introduce and briefly discuss the model, that is, the stochastic delay differential equations that we use in our analysis and in the proof of concept. We then compute and analyze the power spectral density (defined in Sec.~\ref{sec:definitions_psd}) for a system subject to linear time-delayed feedback in Sec.~\ref{sec:linear_results_analytical}. After identifying the signature of time delay 
in linear time-delayed feedback systems, this signature serves as the basis for our general PSD-based approach to the inference of time delay. 
Next, 
after preliminary remarks (Sec.~\ref{sec:prelim_diff}) about the treatment of non-stationary time series, we present numerical results for linear (Sec.~\ref{sec:linear_numerical})
and specific nonlinear feedback forces (Sec.~\ref{sec:nonlinear_numerical_limits}) before demonstrating PSD-based inference for 
fully nonlinear systems in Sec.~\ref{sec:full_nonlin}. Subsequently, we move on to the perturbation-based approach to infer the time delay, where we first discuss the characteristic response of a system with time delay to an external perturbation.
To this end, in Sec.~\ref{sec:characteristic_response} we study the 
response of a deterministic 
linear
system analytically using the method of steps and show that the response 
persists in stochastic systems 
and also carries over to nonlinear systems. We then discuss in Sec.~\ref{sec:cnn} how, using this response, a convolutional neural network (CNN) can be employed to classify such perturbed trajectories of time-delayed stochastic systems.
Subsequently, in Sec.~\ref{sec:perturb_numerical} we present results for time series classification with respect to the delay time by a suitably trained neural network, demonstrating the viability and accuracy of the perturbation based approach. Again, we consider linear feedback forces in Sec.~\ref{sec:cnn_num_lin} as well as more general nonlinear feedback forces (Secs.~\ref{sec:cnn_num_nonlin}-\ref{sec:full_nonlin_network}) in our proof of concept. 
This demonstrates the versatility of the neural network-based classification of time series from delayed stochastic systems.
Finally, we present conclusions and an outlook to potential applications and 
future research directions in Sec.~\ref{sec:conclusions}. 
Details about the simulation method, the inference procedures, data analysis, as well as analytical details, 
supplementary figures and a brief discussion of colored noise systems with corresponding numerical data are given in Appendices.

\section{Model\label{sec:model_basic}} 
To set the stage, 
we consider a one-dimensional system of a 
Brownian particle subject to time-delayed feedback (dFB). 
The dynamics of the Brownian particle is governed by the stochastic delay differential equation (SDDE)
\begin{equation}
    \gamma \frac{dx}{dt} = F_\mathrm{dFB}\left[x(t),x(t-\tau)\right] +\xi(t)\ .
\label{eq:overdamped_Langevin}
\end{equation}
The term $\xi(t)$ 
represents 
Gaussian white noise, with zero mean and correlation function $\langle \xi (t) \xi(t^\prime) \rangle = 2 \gamma k_BT \delta(t-t^\prime)$, where $k_B T$ (with $k_B$ being the Boltzmann constant and $T$ the temperature) is the thermal energy. 
(We here focus on Gaussian white noise for simplicity. However, Appendix~\ref{sec:colored_noise} discusses the applicability and numerical results based on the methods introduced below for systems with exponentially correlated noise.)
The time-delayed feedback force, $F_\mathrm{dFB}[x(t),x(t-\tau)]$, depends on the instantaneous position of the particle $x(t)$ and the earlier position $x(t-\tau)$, where $\tau$ represents the {\it delay} time. 
Instead of an initial condition, as it is required for non-delayed systems, the SDDE is complemented with an initial history function $\Phi(t)$ containing the particle's position for $t\in[-\tau,0]$.
In principle, the SDDE can additionally include the 
force $F_\mathrm{inst}[x(t)]$ that depends {\it only} on the instantaneous position $x(t)$. 
In what follows, we take the particle diameter $\sigma$ as a length scale, and the Brownian 
time  $\tau_\mathrm{B}\equiv\sigma^2/D$ as a time scale (with the bare diffusion constant $D=k_B T/\gamma$).
\subsection{Linear feedback force\label{sec:lfdf}}
For our analytical considerations
we focus on a 
linear time-delayed feedback force $F_\mathrm{dFB}$ of the 
form
\begin{align}
    F_\mathrm{lin}  =-\nabla_{x} V_\mathrm{quad}[x(t),x(t-\tau)] =-[k_a x(t) - k_b x(t-\tau)]\ .
    \label{eq:linear_general}
\end{align}
The corresponding delayed feedback potential, $V_\mathrm{quad}$, can most intuitively be understood in the limit $k_a=k_b$, where it reduces to a harmonic potential centered at the earlier position $x(t-\tau)$.
When $k_a\neq k_b$, the feedback force can be understood as competing instantaneous and delayed forces, where the dominant force is determined by the ratio of parameters $k_a$ and $k_b$. 
Systems subject to such linear feedback forces have been studied extensively both, in deterministic \cite{pyragasContinuousControlChaos1992,kuchler_langevins_1992-1,tsuiControlHigherDimensional2000,yanchukDelayPeriodicity2009} as well as stochastic systems \cite{kuchler_langevins_1992-1,frankStationarySolutionsLinear2001,munakataLinearStochasticSystem2009,munakataEntropyProductionFluctuation2014a,loos_heat_2019}. In the context of deterministic dynamical systems, stability analysis is one of the central points of interest
(see e.g., \cite{yanchukControlUnstableSteady2006,yanchukDelayPeriodicity2009}) and it has, for example, been shown in an early application that time-delayed feedback can be used to stabilize otherwise unstable orbits~\cite{pyragasContinuousControlChaos1992,pyragas_control_1995}. On the other hand, stability is also a central feature of stochastic systems and typically connected to stationary or steady states. For the present class of feedback forces, stability has been assessed for different parameter regimes~\cite{kuchler_langevins_1992-1} as well as in terms of extensions of the model
~\cite{loosMediumEntropyReduction2021}.
%, 
Here, we distinguish two cases in terms of stability.
Firstly, we consider the case when $0<k_b \leq k_a$, 
which we refer to as attractive feedback.
Specifically, attractive feedback is marginally stable for $k_a=k_b$ (attractive harmonic feedback)~\cite{koppHeatProductionStochastic2024,saha_cybloids_2024-1}, implying a constant mean position but a growing positional variance and, thus, a non-stationary time series in the language of time series analysis.  Otherwise, it is stable \cite{kuchler_langevins_1992-1,loosMediumEntropyReduction2021}, i.e., both positional mean and variance are constant in the long time limit.  

Secondly, the case when $k_{a} < 0$ and $k_{b} < 0$ we call repulsive feedback. 
Repulsive feedback exhibits a more complex stability behavior. In the marginally stable case, when $k_a = k_b$ (repulsive harmonic feedback), there exists a threshold value $k_\mathrm{thresh}(\tau) \equiv \gamma/\tau$
in the parameter space (with respect to $|k_a|$ and $|k_b|$) \cite{kopp_persistent_2023-1} above which the feedback forces 
diverge
(as expected for a force derived from an inverse parabola). Below this threshold, when $|k_a| = |k_b| < k_\mathrm{thresh}(\tau)$, 
the system exhibits 
enhanced diffusion \cite{kopp_persistent_2023-1} or even transient super-diffusive behavior \cite{kopp_persistent_2023-1,saha_cybloids_2024-1}, depending on parameter combinations and the distance from the threshold. In the present article,
we study the repulsive system only for parameter combinations below $k_{\rm thresh}$.
Furthermore, when $|k_a| < |k_b| < k_\mathrm{thresh}(\tau)$ and $\tau < \arccos(-k_a/k_b)/\sqrt{k_b^2 - k_a^2}$ the system assumes a stationary state \cite{kuchler_langevins_1992-1}. These conditions are satisfied in our considerations for repulsive feedback.
\subsection{Nonlinear feedback forces\label{sec:nonlinear_analytical}}
Among the broad class of nonlinear feedback forces, we here discuss two examples that have been studied in other contexts. Importantly, linearization of these forces yields the form given in Eq.~\eqref{eq:linear_general}.  

The first force is based on a hyperbolic tangent function (see Ref.~\cite{lichtnerFeedbackcontrolledTransportInteracting2010} for a similar feedback force in the context of feedback-controlled transport and Ref.~\cite{lequyStochasticMotionNonlinear2023} 
for a non-delayed system) that depends on the difference between instantaneous and delayed positions, 
\begin{equation}
    F_\mathrm{tanh} = -A\tanh\big[B\big(x(t)-x(t-\tau)\big)\big]\ .
    \label{eq:tanh_fb_force}
\end{equation}
This feedback force can be linearized
for small magnitudes of $B$ and/or small $\tau$ (in the spirit of a small delay approximation~\cite{guillouzicSmallDelayApproximation1999,guillouzicRateProcessesDelayed2000}), recovering the attractive/repulsive harmonic feedback system discussed above (Sec.~\ref{sec:lfdf}).
In its nonlinear form, it has the advantage of only producing finite feedback forces as opposed to the linear feedback force [Eq.~\eqref{eq:linear_general}] introduced in the previous subsection~\ref{sec:lfdf}. 

Secondly, Gaussian 
time-delayed feedback has been shown to cause peculiar behavior and unusual features  when applied to colloidal particles~\cite{bell-daviesDynamicsColloidalParticle2023}, most significantly persistent motion above a threshold in parameter space \cite{kopp_persistent_2023-1,bell-daviesDynamicsColloidalParticle2023}. 
Here, we consider the force
\begin{align}
    F_\mathrm{Gauss} & =-\nabla_{x} V_{\rm Gauss}[x(t),x(t-\tau)] \label{eq:Gaussian_fb_force}\\
                     & = \frac{a}{b^2}\left[x(t)-x(t-\tau)\right]e^{-\frac{\left[x(t) - x(t-\tau)\right]^2}{2b^2}}\ ,\nonumber
\end{align}
where for $a>0$, $V_\mathrm{Gauss}$ is
a repulsive Gaussian delayed feedback potential 
as studied, for example, in \cite{kopp_persistent_2023-1,koppSpontaneousVelocityAlignment2023,bell-daviesDynamicsColloidalParticle2023}.

To understand the impact of $F_\mathrm{Gauss}$, we first consider Eq.~\eqref{eq:overdamped_Langevin} with the feedback force~\eqref{eq:Gaussian_fb_force} in the absence of a heat bath (i.e., temperature $T = 0$). Above a threshold in the parameter space, i.e., $a\tau/(\gamma b^2)>1$
~\cite{kopp_persistent_2023-1}, the described particle moves at a constant velocity $v_\infty$ in the long time limit~\cite{kopp_persistent_2023-1} 
given a non-zero initial history function.

In the stochastic system, above this threshold, persistent motion is observed at intermediate times, while the feedback leads to enhanced diffusion in the long-time limit.
Further, below the threshold to persistent motion, i.e., $a\tau/(\gamma b^2) < 1$, 
a Brownian particle subject to repulsive 
Gaussian time-delayed feedback exhibits enhanced diffusion \cite{kopp_persistent_2023-1}.

Linearizations of the repulsive
Gaussian feedback force~\eqref{eq:Gaussian_fb_force}  (leading to variants of the two marginally stable linear feedback force systems discussed above [Sec.~\ref{sec:lfdf}, Eq.~\eqref{eq:linear_general} with $k_a=k_b$]) are possible for two cases (when $b$ is fixed).
First, for small displacements (and $a\tau/(\gamma b^2)<1$) we recover a repulsive harmonic feedback force~\cite{kopp_persistent_2023-1} (see Appendix~\ref{sec:gaussian_linearized_small_delay}). Second, for the case where the particle exhibits persistent motion \cite{kopp_persistent_2023-1} (see Appendix~\ref{sec:gaussian_linearized_const_vel}) the linearization can be understood as a co-moving attractive harmonic feedback trap. 

\section{Power spectral density approach -- Methodology\label{sec:psd_analysis}}
In this paper, our main objective is to infer the time delay from noisy data.
In the present section, we introduce and discuss the power spectral density approach.

\subsection{Definitions\label{sec:definitions_psd}}
The power spectral density (PSD) \cite{bartlettPeriodogramAnalysisContinuous1950,blackmanMeasurementPowerSpectra1958} is a well established quantity for the analysis of noisy (measurement noise) or fully stochastic (e.g., Brownian) systems. It is frequently used in experimental and theoretical studies \cite{muenkerAccessingActivityViscoelastic2024,berg-sorensen_power_2004,berg-sorensen_power_2006,sposiniRobustCriterionAnomalous2022,squarcini_spectral_2022-1,krapf_spectral_2019-1,krapf_power_2018,metzler_brownian_2019,mckinleyAnomalousDiffusionGeneralized2018} as well as a wide range of (engineering) applications (e.g., signal analysis \cite{blackmanMeasurementPowerSpectra1958,heFundamentalsMeasurementSignal2022}).  Its main advantages 
are 
accessibility (i.e., it can be directly applied to measurement data and computed using fast algorithms \cite{welchUseFastFourier1967} such as the fast Fourier transform (FFT) \cite{cooleyAlgorithmMachineCalculation1965}) and robustness with respect to noise.

In the context of linear time-delayed feedback, the PSD has previously been defined and computed 
based on 
correlation functions~\cite{guillouzicSmallDelayApproximation1999}  and in the context of 
response theory~\cite{rosinbergStochasticThermodynamicsLangevin2015,rosinbergStochasticThermodynamicsLangevin2017} (see also Appendix~\ref{sec:textbook_psd}).
We here focus on a different definition (often used in signal analysis~\cite{blackmanMeasurementPowerSpectra1958}).
It reads 
\begin{equation}
       S(\omega) 
       \equiv \lim_{\mathcal{T}\to\infty} \frac{1}{\mathcal T}|\tilde Z(\omega)|^2\ ,\label{eq:def_psd}
\end{equation}
where $\mathcal{T}$ is the trajectory length, $\tilde{Z}$ is the Fourier transform of a stationary time series $Z(t)$ (i.e., with mean and variance approaching a constant value) and $\omega$ is the frequency. 
Here, we used the following definition for the Fourier transform:
\begin{align}
    \tilde f(\omega) &= \int_{-\infty}^{+\infty}~dt~e^{-i \omega t}~f(t)\ . 
\end{align}

For a stochastic time series $Z(t)$, 
the power spectral density~\eqref{eq:def_psd} is a fluctuating quantity.
To ensure that the data are suitably smooth for the purpose of inference of the delay time,
%, 
we consider
the noise-averaged power spectral density (from this point we use the acronym PSD as a short hand notation for this quantity). It is defined as
\begin{equation}
       {\rm PSD} \equiv \lim_{\mathcal{T}\to\infty} \frac{1}{\mathcal{T}}\langle|\tilde Z(\omega)|^2\rangle\ ,\label{eq:naPSD_definition}
\end{equation}
where $\langle \ldots \rangle$ 
indicates an ensemble average 
over noise realizations.
To compute the PSD, we 
evaluate the ensemble average
before taking the limit $\mathcal{T}\to\infty$. 
Swapping the average and the limit 
is possible as long as each trajectory is sufficiently long.

\subsection{Linear systems\label{sec:linear_results_analytical}}
Explicit results for the PSD can be obtained for
the linear time-delayed feedback~[Eq.~\eqref{eq:linear_general}] where $k_a$ and $k_b$ are chosen so that the system is stable or marginally stable.  For the repulsive case 
($k_{a,b} < 0$),  we choose the feedback parameters such that we stay below the threshold to instability, i.e.,~$|k_a| \leq |k_b| < k_\mathrm{thresh}(\tau)\equiv \gamma/\tau$. Above the threshold, the system is unstable (resulting in a non-stationary time series due to diverging forces), and, thus, the PSD is not defined.

\begin{figure}[h!]
    \centering
    \includegraphics[width=\columnwidth]{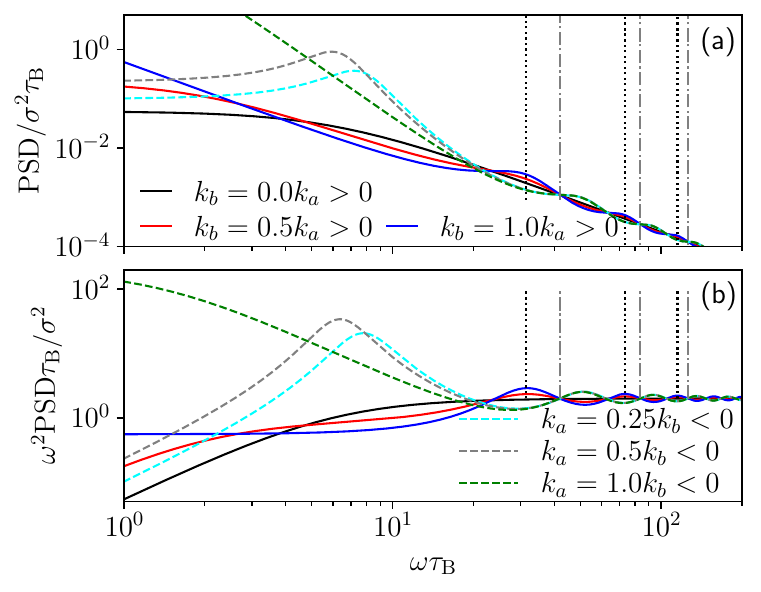}
    \caption{
    Power spectral density (PSD) as a function of the frequency $\omega$ for a Brownian particle subject to linear time-delayed feedback~\eqref{eq:linear_general} at $\tau = 0.15\tau_B$. Solid/dashed lines: Analytical result~\eqref{eq:psd_linear_ana} (panel a) and the PSD multiplied by $\omega^2$ (panel b), for different values of $k_{a,b}$.  Black solid curve represents a harmonically confined Brownian particle without feedback [i.e., $k_b = 0$ in Eq.~\eqref{eq:psd_linear_ana}. Dash-dotted grey vertical lines: Approximate inflection points at $\omega = 2\pi n/\tau$ with $n\in\mathbb{N}$. Dotted black
    vertical lines: Approximate maximum for $k_{a,b}>0$ at $\omega = 3\pi/(2\tau) + 2\pi n/\tau$ for $n>1$, where $n\in\mathbb{N}$.  
    }
    \label{fig:PSD_analytical_var_kb}
\end{figure}

We are interested in the PSD~\eqref{eq:naPSD_definition} with $\tilde{Z}=\tilde{x}$.
The Fourier transform of the linear dynamics [Eq.~\eqref{eq:overdamped_Langevin} with Eq.~\eqref{eq:linear_general}] reads
\begin{align}
    \gamma i\omega \tilde{x}(\omega) &= -(k_a - k_b e^{-i\omega \tau})\tilde{x}(\omega) + \tilde\xi(\omega)\ .
\end{align}
Solving for $\tilde{x}(\omega)$, we get
\begin{align}
    \tilde{x}(\omega) &= \frac{\tilde \xi(\omega)}{\gamma i\omega + k_a - k_b e^{-i\omega \tau}}\ .
    \label{eq:lin_ana_ft}
\end{align}
Substituting Eq.~\eqref{eq:lin_ana_ft} in 
\eqref{eq:naPSD_definition} we obtain 
\begin{align}
        &\langle S(\omega)\rangle_\mathrm{lin} = \label{eq:psd_linear_ana}\\
        &\frac{2\gamma k_B T}{\gamma^2\omega^2-2\gamma k_b\omega\sin(\omega\tau)-2 k_a k_b\cos(\omega\tau)+k_a^2 + k_b^2}\ .
        \nonumber
\end{align}
Figure~\ref{fig:PSD_analytical_var_kb} shows 
analytical results 
\eqref{eq:psd_linear_ana}
for the PSD of the linear system for different $k_a$, $k_b$ and fixed $\tau$.

We first consider the limit
$k_b=0$ in Eq.~\eqref{eq:psd_linear_ana}, i.e., no feedback.
In this case, and if $k_a > 0$, 
the expression takes the shape of the PSD of a Brownian particle in a harmonic trap 
of stiffness $k_a$. 
It follows that 
%. 
$\langle S(\omega)\rangle_\mathrm{lin} =\frac{2\gamma k_B T}{\gamma^2\omega^2 + k_a^2}$, which 
decreases as $1/\omega^2$ in the large-frequency limit (see the non-oscillatory black solid 
curve in Fig.~\ref{fig:PSD_analytical_var_kb}). 
(We note here that for $k_a < 0$, the particle is subject to an instantaneous repulsive harmonic potential, which yields instability.)

When introducing time delay, i.e., $\tau > 0$, the $\tau$-dependent oscillatory terms ($\sin$ and $\cos$) in the denominator of Eq.~\eqref{eq:psd_linear_ana} come into play. As a consequence, the PSD and, even more clearly, the function $\omega^2$PSD [that is, Eq.~\eqref{eq:psd_linear_ana} multiplied by $\omega^2$] show oscillatory behavior, in contrast to the harmonically trapped particle (see Fig.~\ref{fig:PSD_analytical_var_kb}). 
Oscillations induced by time delay are not unexpected. In deterministic systems, this can be an oscillating position while for stochastic systems one would expect oscillations in the mean square displacement~\cite{saha_cybloids_2024-1} or other correlation functions~\cite{munakataLinearStochasticSystem2009,rosinbergStochasticThermodynamicsLangevin2015,rosinbergStochasticThermodynamicsLangevin2017}. Furthermore, for similar linear feedback force systems, we know that at least in stationary states, oscillations can also occur in the PSD (see e.g.~\cite{guillouzicSmallDelayApproximation1999}). Here, we analyze the oscillations in the linear system as a reference for our subsequent investigation of the nonlinear case.

From Fig.~\ref{fig:PSD_analytical_var_kb} one can observe that the minima and maxima for $k_{a,b} > 0$ and $k_{a,b}<0$ are offset by half an oscillation period, where the offset is a consequence of the sign change in the prefactors of the trigonometric functions [Eq.~\eqref{eq:psd_linear_ana}].

Importantly, the specific choice of the parameters $k_{a,b}$, 
only influences the small frequency behavior (see Appendix~\ref{sec:small_freq} for an 
analytical 
consideration of this limit and Appendix~\ref{sec:marginally_stable_app} for the special case of a harmonic feedback system) but does not affect the oscillatory part of the PSD 
on which we concentrate from now on.

\begin{figure}[h!]
    \centering
    \includegraphics[width=\columnwidth]{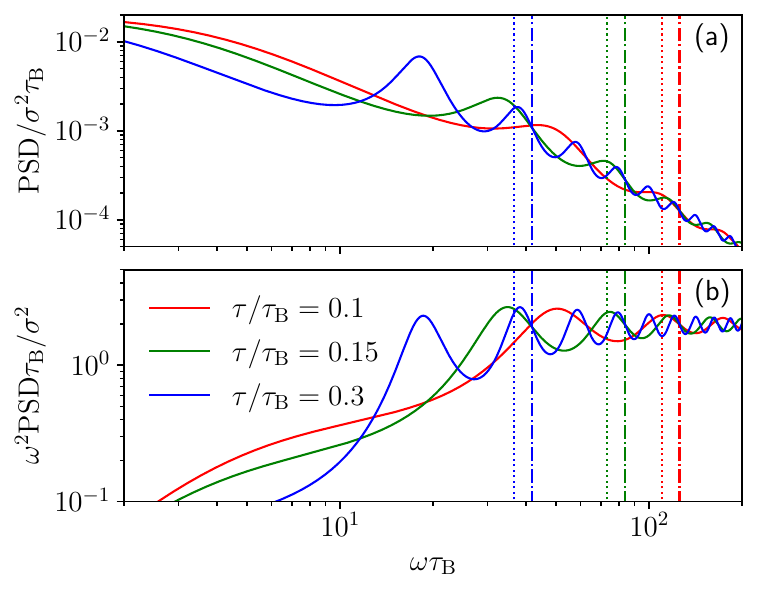}
    \caption{
    Power spectral density (PSD) as a function of the frequency $\omega$ for a Brownian particle subject to linear time-delayed feedback~\eqref{eq:linear_general} for $k_b = 0.5 k_a>0$. Solid lines: Analytical results~\eqref{eq:psd_linear_ana} (panel a) and the PSD multiplied by $\omega^2$ (panel b) for different values of the delay time $\tau$. 
    Dash-dotted vertical lines: Approximate inflection points at $\omega = 2\pi n/\tau$ with $n\in\mathbb{N}$. Dotted vertical lines: Approximate maximum at $\omega = 3\pi/(2\tau) + 2\pi n/\tau$ for  $n>1$, where $n\in\mathbb{N}$.
    }
    \label{fig:PSD_analytical_var_tau}
\end{figure}

To illustrate the $\tau$-dependence of the oscillations, we present the PSD and the function $\omega^2\mathrm{PSD}$ at different values of the delay time (color coded) in Fig.~\ref{fig:PSD_analytical_var_tau} for a system with $k_a>k_b > 0$.
While the term $2k_ak_b\cos(\omega\tau)$ in Eq.~\eqref{eq:psd_linear_ana} has a maximum magnitude of $|2k_ak_b|$, the magnitude of $2\gamma k_b\omega\sin(\omega\tau)$ grows with $\omega$. Therefore, as $\omega$ grows, specifically when 
$\omega > k_a/\gamma$, the term $2\gamma k_b\omega\sin(\omega\tau)$ 
[in Eq.~\eqref{eq:psd_linear_ana}] becomes dominant. 
Indeed, we observe from Fig.~\ref{fig:PSD_analytical_var_tau} that the value of the delay time determines both, the location of the maxima and minima 
and the oscillation period. The first minimum/maximum after the sine term becomes dominant is the first relevant maximum/minimum for attractive/repulsive feedback. It is located at approximately $\omega=3\pi/(2\tau)$. For $\omega \gg k_a/\gamma$, the oscillation period [due to the periodicity of the $\sin$ and $\cos$ terms in Eq.~\eqref{eq:psd_linear_ana}] is given by $\Delta\omega = 2\pi/\tau$ 
in frequency space (dotted and dash-dotted vertical lines of the same color indicating the approximate location of the maxima and inflection points, respectively). We refer to this behavior, i.e. oscillations with period $\Delta\omega = 2\pi/\tau$ occurring at large $\omega$, as the \textit{signature of time delay}.

For longer delay times, we observe from Fig.~\ref{fig:PSD_analytical_var_tau} a shift of the delay-time dependent positions of the minima and maxima 
to smaller frequencies and a shorter oscillation period, consistent with the prediction.
Thus, 
the specific oscillatory behavior and delay-time dependent oscillation period $\Delta\omega = 2\pi/\tau$ 
can, in principle, be used to infer time delay without exact knowledge about the other parameters in the system or fitting simulation data with the analytical result.

\begin{figure}[]
    \centering
    \includegraphics[width=\columnwidth]{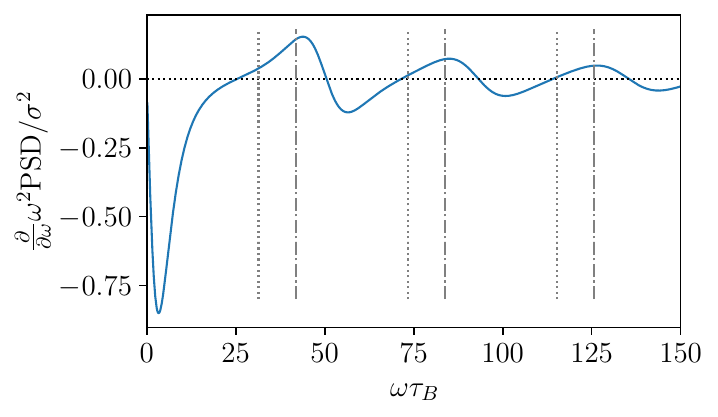}
    \caption{
    The derivative of PSD~\eqref{eq:psd_linear_ana} [i.e., $\partial/\partial\omega [\omega^2{\rm PSD}(\omega)$] as a function of $\omega$. Here, $k_{a,b} > 0$ and $k_a\neq k_b$. Dotted horizontal line: Guide for the eye to indicate the zero-crossings of the blue solid curve. Dash-dotted vertical line: Inflection points of $\omega^2{\rm PSD}$ at $\omega = 2\pi n/\tau$, for
    $n \in \mathbb{N}$. Dotted vertical line: maxima of $\omega^2{\rm PSD}$ at $\omega = 3 \pi /(2\tau) + 2\pi n/\tau$.
    }
    \label{fig:der_omegasqpsd}
\end{figure}

To extract the positions of the maxima and minima in the frequency space,
it is instructive to consider the derivative 
of $\omega^2 {\rm PSD}$
with respect to $\omega$. This is depicted in Fig.~\ref{fig:der_omegasqpsd} for the case $k_{a,b} > 0$. The zero crossings correspond to the positions of the minima and maxima.
For $k_{a,b}>0$ the maxima of $\omega^2\mathrm{PSD}$, which are roots of $\partial_\omega [\omega^2 {\rm PSD}]$, 
are located at $\omega \approx  3\pi/(2\tau) + 2\pi n/\tau$ (dotted vertical lines in Fig.~\ref{fig:der_omegasqpsd}) while the minima are offset by $\pi/\tau$. 
The inflection points of $\omega^2\mathrm{PSD}$ (maxima in Fig.~\ref{fig:der_omegasqpsd}) are located at $\omega \approx 2\pi n/\tau$, $n \in \mathbb{N}^+$ (indicated by dash-dotted vertical lines). We observe from Fig.~\ref{fig:der_omegasqpsd}, specifically from the second set of vertical lines at $\omega\tau_B\sim75$, that 
the approximation is in good agreement with the actual positions of the roots and maxima of 
$\partial_\omega [\omega^2 {\rm PSD}]$, while for the first zero crossing and maximum there is still a significant deviation. At those frequencies, the term containing $\cos(\omega\tau)$ is not negligible.

\section{Power spectral density approach -- numerical results\label{sec:psd_results}}
We now discuss numerical results (based on trajectories from BD simulations, see Appendix~\ref{sec:simulation_method} for details) for the PSD for various types of SDDEs. 
We first benchmark the numerics by considering strictly linear systems. 
We then proceed by considering the specific nonlinear systems introduced in Sec.~\ref{sec:nonlinear_analytical}, where linear approximations recover harmonic attractive/repulsive feedback systems. This allows us to obtain first insights into the impact of nonlinearity.
Finally, we move on to more general nonlinear systems
to see how the oscillations can be used to infer the delay time without explicit knowledge of the functional form of the underlying forces.

\subsection{Preliminary remarks\label{sec:prelim_diff}}
For the computation of the PSD it is crucial that the underlying time series is stationary (see Appendix~\ref{sec:differencing} for detailed discussion on this topic), which means that for the mean and variance of $Z(t)$ in Eq.~\eqref{eq:naPSD_definition} 
have to approach constants as a function of time. Already a simple passive Brownian particle 
does not fulfill 
this requirement, since its positional variance grows in time.
Similarly, the marginally stable linear feedback system effectively exhibits diffusive behavior in the long time limit. To overcome this problem, we apply the 
differencing method (see Appendix~\ref{sec:differencing}), i.e., we compute the PSD of the positional difference time series $\{\bar{x}(t)\}\equiv \{x(t+dt) - \{x(t)\}$ for small $dt$, instead of the position itself.
As a rough analogy, this method can be understood as `taking the discretized derivative' of the positional data with respect to time. After applying the Fourier transform and dividing by $dt$ as well as by the frequency (recall that derivatives become algebraic factors under the Fourier transform), this leaves us with the correct Fourier transform of the position, which can then be plugged into 
Eq.~\eqref{eq:naPSD_definition}. Not applying this technique would lead to a significant deviation between numerical results and analytical prediction, which become particularly severe for the (non-stationary) repulsive feedback case, as shown in Appendix~\ref{sec:differencing}, Fig.~\ref{fig:differencing_method_numerical}. 
For consistency, we apply the same procedure to all of our data, even for systems that are stationary in the long time limit.
\begin{figure}
    \centering
    \includegraphics[width=\columnwidth]{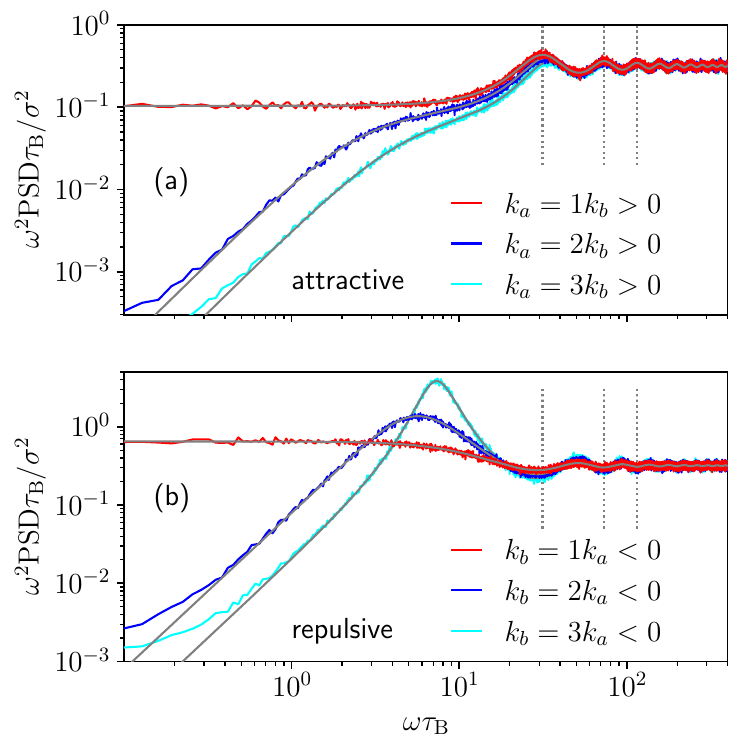}
    \caption{
    Comparison of analytical power spectral density (PSD) (solid grey curves)~\eqref{eq:psd_linear_ana} with numerically simulated noise-averaged PSD as a function of frequency $\omega$ for the linear system~\eqref{eq:linear_general} at $\tau = 0.15 \tau_B$. Panel a): $k_{a,b}>0$. Panel b): $k_{a,b}<0$, where we take $|k_b| < k_\mathrm{thresh}(\tau)$ to ensure finite feedback forces. Vertical dotted grey lines: Approximate analytical locations of maxima/minima for attractive/repulsive feedback. Numerical results are obtained by noise averaging over 400 noise realizations.  
    }
    \label{fig:linear_system_general}
\end{figure}
\begin{figure}[h!]
    \centering
    \includegraphics[width=\columnwidth]{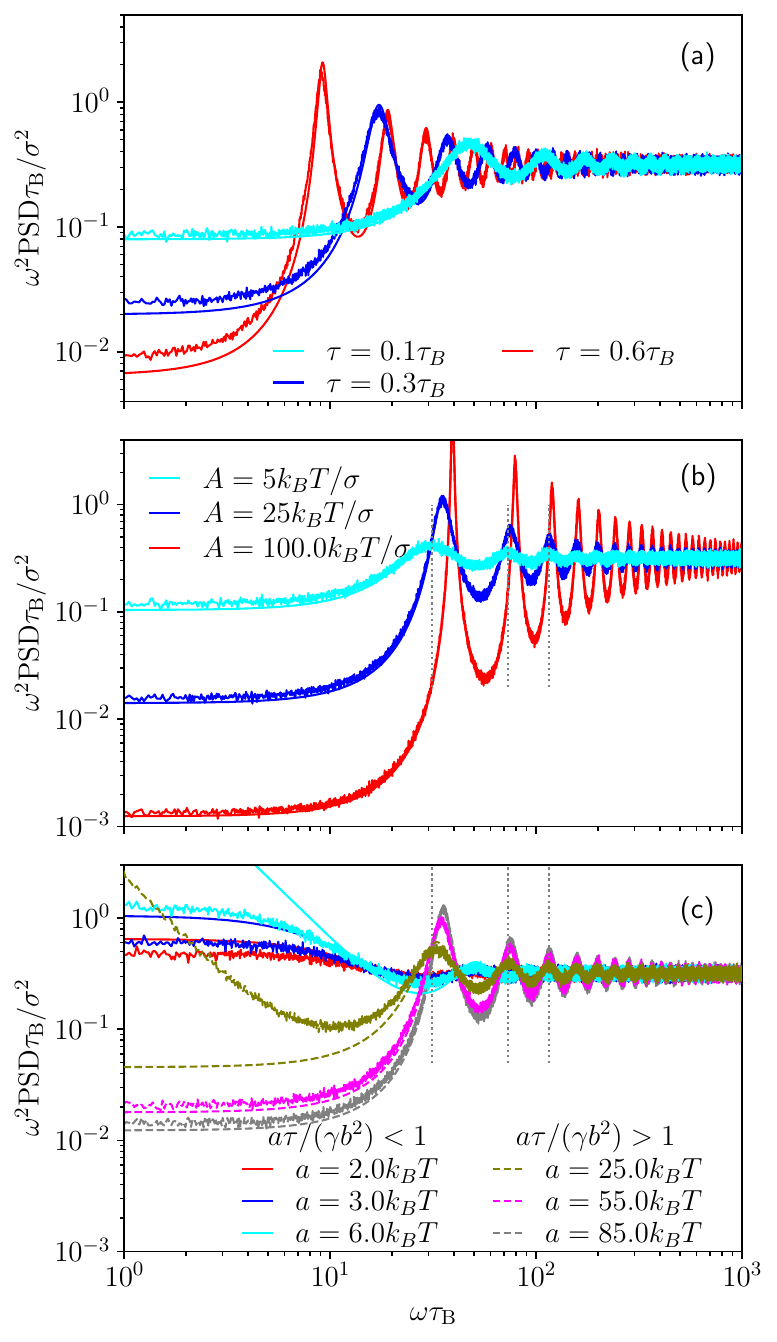}
    \caption{
    Comparison of numerically simulated (colored points, obtained from averaging over 400 noise realizations) power spectral density (PSD) of nonlinear systems with the linear approximation (solid/dashed curves). Panel (a): $\tanh$ feedback force for various $\tau$; solid curves: Linear approximation with $k_a = k_b = 10k_\mathrm{B}T/\sigma$. 
    Panel (b): $\tanh$ feedback force~\eqref{eq:tanh_fb_force} for different $A$; solid curves: Linear approximation with corresponding $k_a = k_b > 0$. 
    Panel (c): Gaussian feedback force~\eqref{eq:Gaussian_fb_force}; solid (dashed) curves:  Analytical result for the linear approximation with repulsive harmonic feedback ($k_a = k_b < 0$) and modified attractive harmonic feedback for the constant velocity state [see Appendix \ref{sec:gaussian_linearized_const_vel}, Eq.~\eqref{eq:linear_const_vel} for more details]. Panel (a): $B=1/\sigma$. Panels (b) and (c): $\tau = 0.15\tau_\mathrm{B}$. Vertical grey dotted lines: Approximate locations of maxima/minima according to the analytical result for the linear system in panels (b) and (c). 
    }
    \label{fig:nonlinear_numerical}
\end{figure}

\subsection{Linear systems\label{sec:linear_numerical}}
Figure~\ref{fig:linear_system_general}  
presents numerical results based on Eq.~\eqref{eq:overdamped_Langevin} with a feedback force as in Eq.~\eqref{eq:linear_general} in the absence of additional instantaneous forces. As before, we distinguish attractive feedback (Fig.~\ref{fig:linear_system_general}a) 
and repulsive feedback (Fig.~\ref{fig:linear_system_general}b).
In the regime $\omega\tau_B \gtrsim 1$, we find almost perfect agreement with the analytical prediction for all cases considered. 
Only at small frequencies and $k_a \neq k_b$ 
we observe a deviation from the analytical predictions. The deviation can be shifted to smaller frequencies by using longer trajectories. We therefore understand this behavior as an effect connected to the finite length of the trajectory used to compute the PSD.

\subsection{Nonlinear feedback systems and their linear approximations\label{sec:nonlinear_numerical_limits}}
Now we turn towards the nonlinear systems based on the hyperbolic tangent [Eq.~\eqref{eq:tanh_fb_force}]
and Gaussian feedback forces [Eq.~\eqref{eq:Gaussian_fb_force}] 
that 
reduce to 
attractive/repulsive harmonic feedback systems (discussed in Sec.~\ref{sec:linear_results_analytical}) in certain limits.
Our goal here is to investigate whether the oscillatory behavior persists in nonlinear systems and under what conditions deviations from linear behavior are observed in $\omega^2$PSD.
Numerical data for $\omega^2\mathrm{PSD}$ 
are presented in Fig.~\ref{fig:nonlinear_numerical}.
Solid lines represent 
analytical results
obtained for the different limits discussed in Sec.~\ref{sec:nonlinear_analytical} and Appendix~\ref{sec:gaussian_linearized_const_vel}.

First, we 
consider
the hyperbolic tangent feedback force~\eqref{eq:tanh_fb_force}.
Figure~\ref{fig:nonlinear_numerical}a and b
present numerical data of the PSD and the 
corresponding analytical approximations based on the attractive harmonic feedback system [Eq.~\eqref{eq:linear_general} with $k_a = k_b > 0$], for different $\tau$ and $A$ respectively.
For both cases, we observe good agreement between numerical data and the analytical prediction for larger $\omega$ (i.e., $\omega\tau_\mathrm{B}\gtrsim 10$ in Fig.~\ref{fig:nonlinear_numerical}), especially with respect to the oscillations.
For small $\omega$, we observe noticeable deviations from the analytical linear prediction. 
These deviations grow as $\tau$ is increased or $A$ is decreased.
This is because we are approaching the limit where the linear approximation becomes invalid.
Similar to linear systems (Fig.~\ref{fig:PSD_analytical_var_tau}), we find 
for this nonlinear system that the delay time determines the oscillation period, as well as the location of the minima and maxima.

Moving on to the system subject to Gaussian time-delayed feedback~[Eq.~\eqref{eq:Gaussian_fb_force}], 
we have to distinguish two limits based on the deterministic threshold, i.e., $a\tau/(\gamma b^2)=1$. 
Far below this threshold (in particular $a = 2k_B T$ and $a = 3k_B T$), in the limit of small displacements  
the feedback force can 
be approximated as a repulsive harmonic feedback \cite{kopp_persistent_2023-1}. On the other hand, for feedback parameters far beyond the threshold to persistent motion ($A = 55k_BT$ and $A=85k_BT$), the particle can be considered as moving at an approximately constant velocity while being subject to attractive harmonic feedback \cite{kopp_persistent_2023-1} (see Appendix \ref{sec:gaussian_linearized_const_vel}).
From Fig.~\ref{fig:nonlinear_numerical}c, far above and below the threshold, we see agreement between the data obtained by numerical simulations of the full nonlinear problem and the 
expression Eq.~\eqref{eq:psd_linear_ana} and the result based on Eq.~\eqref{eq:lin_ana_const_vel}, respectively. 
Particularly, we observe oscillations with the same characteristics as in the linear system. 
Only at low frequencies, i.e., before the onset of oscillatory behavior, do the simulation data deviate from the analytical approximation. 

Interestingly, as we approach the threshold, $a\tau/(\gamma b^2) = 1$, from above or below ($a=6k_B T$ and $a=25 k_B T$) we see a significant deviation from the analytical prediction at small $\omega$, indicating that the simple linear approximation is no longer valid but 
a crossover behavior becomes relevant.
However, the locations of the maxima and minima 
remain unaffected, from which we interpret that the signature of time delay persists even outside the linear regime.

From the results discussed so far, we conclude that the characteristic oscillations with period $\Delta \omega = 2\pi/\tau$ are robust against variation of the feedback force, even in the nonlinear regime.
\subsection{General nonlinear time-delayed feedback\label{sec:full_nonlin}}
We now turn our attention to general nonlinear feedback forces and nonlinear instantaneous forces.

While for the previously discussed systems, in principle, a fit with the exact analytical expression 
is a way to extract the delay time, for a more general functional form there may not be a direct correspondence. 

However, we can make use of the observation for the linear system that, at large $\omega$, the minima and maxima are $\Delta \omega \approx 2\pi/\tau$ apart. 
We 
hypothesize
that oscillations with this oscillation period can also be found in the function $\omega^2$PSD of nonlinear systems, i.e., that the signature of time delay carries over from linear to more general nonlinear systems.
On this basis, we can design a numerical procedure based on a polynomial fit in order to extract the positions of the minima and maxima in frequency space from numerical data and subsequently compute an approximate inferred delay time (for details of the procedure, see Appendix~\ref{sec:psd_procedure}). 

In the following, we present numerical results where we apply this procedure
to three combinations of nonlinear feedback forces and nonlinear instantaneous forces. 
\begin{figure}[h]
    \centering
    \includegraphics[width=\columnwidth,trim={0 .58cm 0 0},clip]{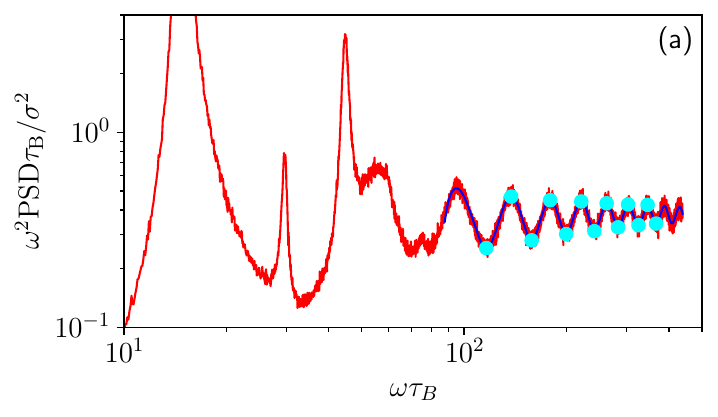}
    \includegraphics[width=\columnwidth,trim={0 .58cm 0 0},clip]{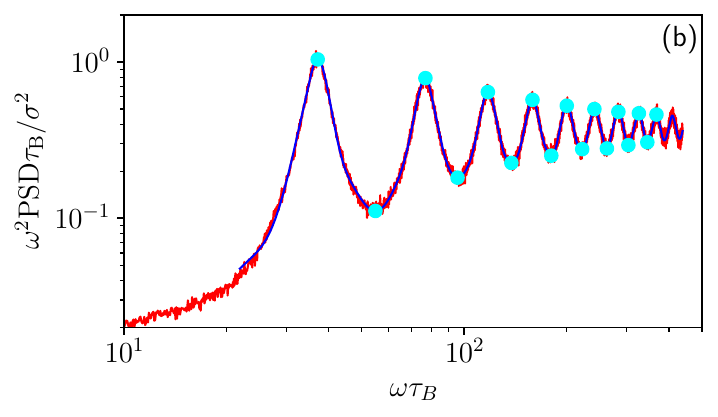}
    \includegraphics[width=\columnwidth]{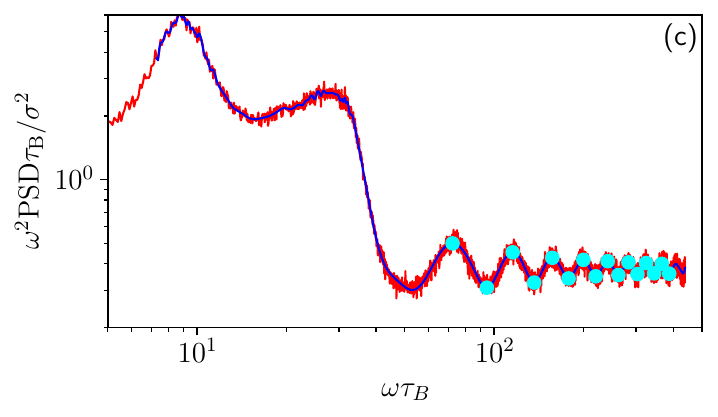}
    \caption{
    Numerically simulated PSD (red) as a function of $\omega$ for non-linear systems~\eqref{eq:dw_linear}, (Panel a)~\eqref{eq:dw_nonlinear} (Panel b), and ~\eqref{eq:tilted_washboard_nonlinear} (Panel c).  
    In each panel: Polynomial fit (blue) using the quartic splines and automatically extracted positions (cyan circles) of maxima and minima using the roots of the derivative of the polynomial (see Appendix \ref{sec:psd_procedure}). See Table~\ref{tab:inferred_delay_times} for the corresponding inference results.
    }
    \label{fig:interpolation}
\end{figure}

The corresponding SDDEs read 
\begin{align}
    \gamma \dot{x}(t) &= -ax^3(t) + bx(t) -cx(t-\tau) + \xi(t)\ ,
    \label{eq:dw_linear} \\
    \gamma \dot{x}(t) &= -ax^3(t) + bx(t) -cx^2(t-\tau) + \xi(t)\ , 
    \label{eq:dw_nonlinear}\\
    \gamma \dot{x}(t) &= -a \cos[x(t)] - b + c\cos[x(t-\tau)] + \xi(t)\ .
    \label{eq:tilted_washboard_nonlinear}
\end{align}
Figure~\ref{fig:interpolation} 
presents exemplary fits of the resulting function $\omega^2$PSD including the obtained positions of the minima and maxima, that were used to infer the delay time. 
The extracted locations of the minima and maxima are in good agreement with the data as shown in Fig.~\ref{fig:interpolation}, where the extracted locations are marked as circles.

Table~\ref{tab:inferred_delay_times} 
summarizes the inference results, where we compare true $\tau$ and inferred delay times $\tau_{\rm inferred}$ for simulation data based on Eqs.~\eqref{eq:dw_linear}, 
~\eqref{eq:dw_nonlinear} and 
~\eqref{eq:tilted_washboard_nonlinear} for three different delay times.
\begin{table}[h]
    \centering
    \begin{tabular}{|c|c||c|c||c|c|}
        \hline
        (a) & Eq.~\eqref{eq:dw_linear} & (b) & Eq.~\eqref{eq:dw_nonlinear} & (c) & Eq.~\eqref{eq:tilted_washboard_nonlinear}\\
        \hline
        $\tau$ & $\tau_\mathrm{inferred}$ & $\tau$ & $\tau_\mathrm{inferred}$ & $\tau$ & $\tau_\mathrm{inferred}$\\
        \hline
        \hline
        $0.1$ & $0.100 \pm  0.002$ & $0.1$ & $0.101 \pm 0.001 $  & $0.1$ & $ 0.100\pm0.005$ \\
        \hline
        $0.15$ & $0.151 \pm 0.001$ & $0.15$ & $0.152 \pm 0.003$ & $0.15$ & $0.149 \pm 0.003$ \\
        \hline
        $0.3$ & $0.303 \pm 0.014$  & $0.3$ & $0.301 \pm 0.009$  & $0.3$ & $ 0.300\pm 0.007$  \\
        \hline
    \end{tabular}
    \caption{
    Delay times $\tau_\mathrm{inferred}$ in units of $\tau_\mathrm{B}$ inferred from PSD with one standard deviation as uncertainty: (a) Data corresponding to Eq.~\eqref{eq:dw_linear} with $a = 10k_\mathrm{B}T/\sigma^4$ and $b=c=10k_\mathrm{B}T/\sigma^2$; (b) Data corresponding to Eq.~\eqref{eq:dw_nonlinear} with $a = 10k_\mathrm{B}T/\sigma^4$, $ b= 10k_\mathrm{B}T/\sigma^2 $ and $c=20k_\mathrm{B}T/\sigma^3$; Data corresponding to Eq.~\eqref{eq:tilted_washboard_nonlinear} with $a = 10k_\mathrm{B}T/\sigma$, $ b = 5k_\mathrm{B}T/\sigma$ and $c=20k_\mathrm{B}T/\sigma$. 
    }
    \label{tab:inferred_delay_times}
\end{table}

Overall, we find robust and accurate inference results for various delay times with an uncertainty of the order of one percent or less. The accuracy of the inference is similar for the different linear and nonlinear feedback and instantaneous forces. Crucially, for the described procedure, knowing the functional form of the forces (instantaneous and delayed) is not required. Therefore, this procedure can be applied to a wide range of systems with discrete time delay.
Furthermore, the procedure is not limited to Gaussian white noise systems only. In Appendix~\ref{sec:colored_noise} we present numerical results for systems with exponentially correlated noise (see Fig.~\ref{fig:PSD_colored_noise_inference} and Table~\ref{tab:inferred_delay_col_noise_fig}).

\section{Perturbation and deep learning-based approach -- Methodology\label{sec:perturb_method}}

Up to this point, we have demonstrated that 
the power spectral density (PSD) features a characteristic 
signature of time delay 
for a broad class of feedback forces (linear and nonlinear) in stochastic systems. This allowed us to infer time delay 
even if the exact functional form of the underlying 
forces is unknown. As a drawback, this methodology requires a large number of 
long trajectories. This can be a challenge in experiments, where one may only have fewer and shorter trajectories at ones disposal.
\begin{figure}[]
    \centering
    \includegraphics[width=0.9\columnwidth]{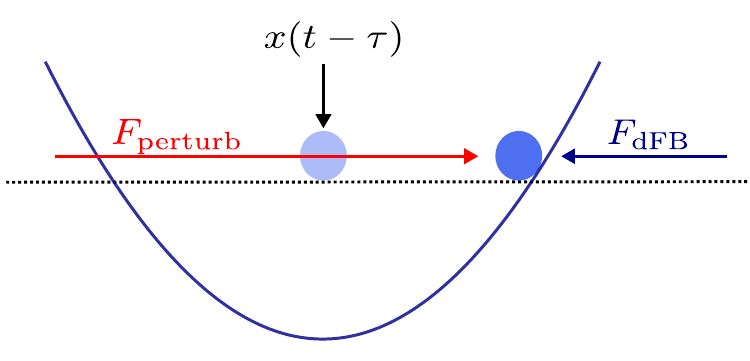}
    \caption{
    Schematic depiction: A particle (dark blue circle), subject to attractive harmonic time-delayed feedback with feedback force~\eqref{eq:linear_general} for $k_a=k_b>0$ in one dimension. The feedback potential is centered at the delayed position $x(t-\tau)$ (i.e., at the fainted blue circle). A perturbing force $F_\mathrm{perturb}$ is applied [Eq.~\eqref{eq:det_delay_perturb}].
    }
    \label{fig:schematics_perturbation}
\end{figure}

To circumvent this issue, we here propose an alternative procedure to infer time delay.
This procedure is based on the observation of the response of the system under the influence of a perturbative force. 
Broadly speaking, 
for a delay-driven system subjected to a perturbation, 
we expect the system's
response to occur 
not instantaneously but rather with some delay,
as we shall see explicitly in the following Subsec.~\ref{sec:characteristic_response}.
Assuming the perturbation to be a force that is applied instantaneously and that is sufficiently strong (i.e., strong compared to noise, feedback forces and instantaneous forces), we 
expect a distinct delayed reaction, a \textit{characteristic response}, as we explain in detail in the following.

\subsection{Characteristic response of a system with time delay to a perturbation\label{sec:characteristic_response}}
To develop an analytical understanding of the characteristics of the response, we take one step back and consider a deterministic system (i.e., the limit $
T \to 0$) with
linear time-delayed feedback force 
and an 
additional perturbative force $F_\mathrm{perturb}(t)$. 
Its equation of motion reads 
\begin{equation}
    \gamma \dot{x} = -[k_a x(t) - k_b x(t-\tau)] + F_\mathrm{perturb}(t)\ .
    \label{eq:det_delay_perturb}
\end{equation}

Intuitively, in the limit $k_a = k_b > 0$ (attractive harmonic), the linear feedback force can be understood as the consequence of a delayed harmonic potential centered at $x(t-\tau)$. 
Figure~\ref{fig:schematics_perturbation} shows 
its schematic representation.
We assume that the particle was 
at rest at $x(t) = 0$ in the time interval $t\in[-\tau,0]$.
We then apply a perturbative force, $F_{\rm perturb}$, starting at $t=0$ for a duration $T_\mathrm{perturb} > \tau$.
The particle will first be confined to a harmonic potential centered at $x(t-\tau) = 0$. 
Following dynamics~\eqref{eq:det_delay_perturb}
after $t=\tau$, the delayed position $x(t-\tau)$ 
(and hence the location of the harmonic trap)
will also start to change, 
resulting in 
a 
change in dynamical behavior. 
% %

\begin{figure*}[]
    \centering
    \includegraphics[width=\columnwidth]{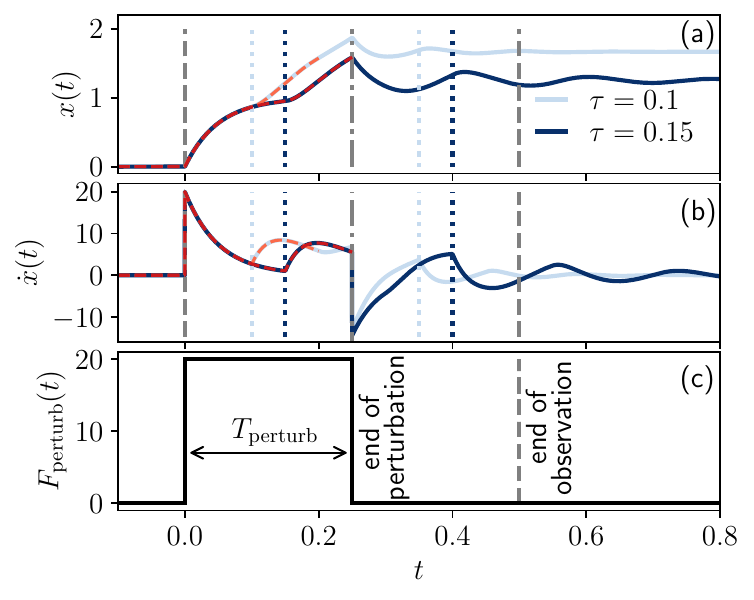}
    \includegraphics[width=\columnwidth]{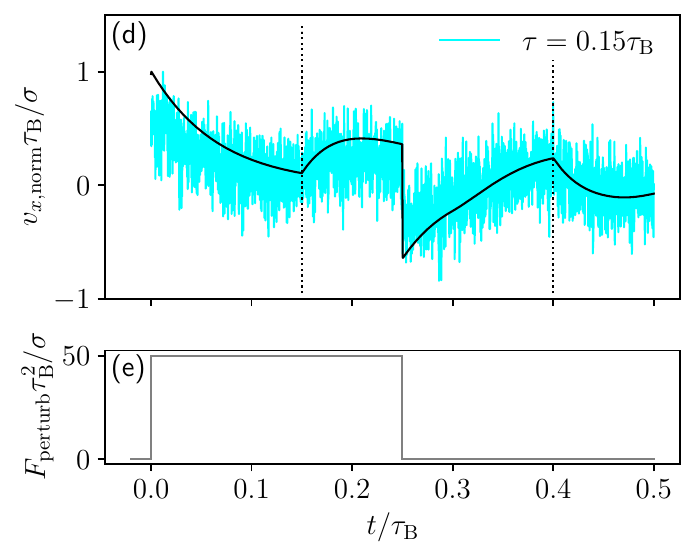}
    \caption{
    Effect of a perturbation on deterministic (a-c) and stochastic systems (d-e). Panels (a-c): The particle subject to attractive harmonic time-delayed feedback and external perturbation~\eqref{eq:det_delay_perturb}. Red dashed lines: Analytical solutions for position~\eqref{eq:mos_expressions} [panel (a)] and velocity~\eqref{eq:der_mos_expressions} [panel (b)] up to $t=2\tau$ for $\tau = 0.1$ and up to the end of the perturbation for $\tau = 0.15$. Dash-dotted vertical line indicates end of perturbation; dashed grey vertical lines represent the beginning and end of observation; vertical dotted lines (color-coded) serve as a guide for the eye at one delay time $\tau$ after the onset, and then, after the end of the perturbation.
    Panel (d) shows the discretized `velocity' $v_{x,\mathrm{norm}}$ after noise averaging over $25$ realizations, applying a simple moving average over three time steps and normalization with the maximum velocity $v_\mathrm{max}$ as a function of time during the observation window; the deterministic velocity (black, from panel b) is shown as a reference. Dotted vertical lines serve as a guide for the eye at $t = \tau$ and $t = T_\mathrm{perturb}+\tau$.
    Panels (c and e): The perturbing force $F_\mathrm{perturb}$ as a function of time. The system is perturbed for a duration of
    $T_\mathrm{perturb} = 0.25 > \tau$ (a-c) and $T_\mathrm{perturb} = 0.25\tau_B > \tau$ (d-e), and the observation continues for another $T_\mathrm{perturb}$.
    }\label{fig:det_perturb_response}
\end{figure*}
To make analytical progress,
we now assume a perturbation protocol given by a step function, i.e., $F_\mathrm{perturb}(t) = F\Theta(t)$, where $\Theta(t)$ is the Heaviside function. 
We 
calculate the dynamics resulting from Eq.~\eqref{eq:det_delay_perturb} using 
the so-called \textit{method of steps} 
\cite{erneuxAppliedDelayDifferential2009}.
This method can be used to obtain an analytical solution 
in 
intervals of one delay time iteratively,
provided that the history function $\Phi(t)$ containing the position of the particles in the previous time interval
is known, i.e., $\Phi(t) = x(t)$ for $t\in[t_0-\tau, t_0]$.
Formally, the 
piecewise solution 
as a function of time during 
the interval $[t_0,t_0+\tau]$ 
reads
\begin{widetext}
\begin{align}
    x_{[t_0,t_0+\tau]}(t) =& \Phi(t_0) e^{-\frac{k_a}{\gamma}(t-t_0)}  + \frac{k_b}{\gamma}\int_{t_0}^t~dt' \Phi(t^\prime - \tau)e^{-\frac{k_a}{\gamma}(t-t^\prime)}   + \frac{F}{\gamma}\int_{t_0}^t~dt'~\Theta(t') e^{-\frac{k_a}{\gamma}(t-t^\prime)}\ . \label{eq:mos-soln}
\end{align}
\end{widetext}
Similarly, the particle's position $x(t)$ can be evaluated for 
$t \geq t_0+\tau$
iteratively. However, for our analytical considerations, we focus on the dynamical behavior during the first two delay times $\tau$ after switching on the perturbation 
at $t = 0$.

We assume that 
the initial history function represents a particle at rest in position $x = 0$,
i.e., we choose $\Phi(t) = x(t) = 0$ for $t\in[-\tau,0]$ (which can easily be realized by pinning the particle). 
The analytical expressions for the position as a function of time during the time interval $[-\tau,2\tau]$ following Eq.~\eqref{eq:mos-soln} then reads 
\begin{widetext}
\begin{equation}
   x(t) = 
   \begin{cases} 
      0\ ,&-\tau < t < 0; \\
      \frac{F}{k_a} \left(1-e^{-k_a t}\right)\ ,& 0\leq t < \tau; \\
      \frac{F}{k_a}\left[2-e^{-\frac{k_a t}{\gamma}}-e^{-\frac{k_a}{\gamma}(t-\tau)}- \frac{k_b}{\gamma} e^{\frac{k_a\tau}{\gamma}}(t-\tau)\right],& \tau \leq t < 2\tau\ . 
   \end{cases}
   \label{eq:mos_expressions}
\end{equation}
For reasons that will become clear later, 
we also take
the time derivative of this piecewise solution~\eqref{eq:mos_expressions}, yielding 
\begin{equation}
   \dot{x}(t) = 
   \begin{cases} 
      0\ ,&-\tau < t < 0; \\
      Fe^{-k_a t}, & 0\leq t < \tau; \\
      \frac{F}{k_a}\left[\frac{k_a}{\gamma} e^{-\frac{k_a t}{\gamma}}+ \frac{k_a}{\gamma} e^{-\frac{k_a}{\gamma}(t-\tau)}- \frac{k_b}{\gamma} e^{\frac{k_a\tau}{\gamma}}\right], & 
      \tau \leq t < 2\tau\ . 
   \end{cases}
   \label{eq:der_mos_expressions}
\end{equation}
\end{widetext}
One first observation from Eqs.~\eqref{eq:mos_expressions} and~\eqref{eq:der_mos_expressions} is 
that the 
piecewise solution~\eqref{eq:mos_expressions} and also its derivative~\eqref{eq:der_mos_expressions} are proportional to the perturbative force $F$. Thus, in the absence of a perturbative force, the particle stays at rest at the initial position $x(0)=0$, as expected in the deterministic case.

Second, Eqs.~\eqref{eq:mos_expressions} and~\eqref{eq:der_mos_expressions} directly reveal the role of the delay time, $\tau$. Indeed, in the first interval $[0,\tau]$, the system behaves as if there was no delay at all. This changes when we consider the regime $t\geq \tau$: 
While the first term encodes the systems instantaneous reaction to the applied perturbation, the second and third term can be understood as the time-delayed response.
Those terms can lead to a significant change in dynamical behavior, 
where for each specific $\tau$ the first change in dynamical behavior occurs immediately at $t=\tau$. This means (given a zero history) that from this point on memory effects become important.

Figure~\ref{fig:det_perturb_response} presents corresponding exemplary numerical data including the analytical results from Eqs.~\eqref{eq:mos_expressions} and \eqref{eq:der_mos_expressions}. We 
perturb the system by applying a constant force to the particle 
starting at $t = 0$ for a duration $T_\mathrm{perturb}$, which has to be, 
at least, 
longer than one delay time
(i.e., $\tau < T_\mathrm{perturb}$) 
and observe the system.
The observation continues after the perturbation is set to zero at $t = T_\mathrm{perturb}$ for a further duration of $T_\mathrm{perturb}$ 
(Fig.~\ref{fig:det_perturb_response}c).
We see from Fig.~\ref{fig:det_perturb_response} that after applying the perturbation, but also after the perturbation is set to zero, there is a characteristic change in dynamical behavior after one $\tau$ [indicated by color coded vertical dotted lines in Fig.~\ref{fig:det_perturb_response}a and b].
This change consists in `kinks' and other deviations from a system without delay.
We also find that the change is
most distinct in $\dot{x}$ (Fig.~\ref{fig:det_perturb_response}b),
even though this is not immediately obvious from the analytical expression~\eqref{eq:der_mos_expressions}.
Importantly, such a
characteristic response is not 
limited to linear systems but carries over to more complex nonlinear instantaneous and delayed forces; see 
Fig.~\ref{fig:response nonlinear forces deterministic} (Appendix) for exemplary data.

But how can this knowledge be exploited in a noisy Brownian system with time delay? Does the characteristic response described before carry over to noisy systems? 
To this end, in Fig.~\ref{fig:det_perturb_response}d
we show how a linear stochastic system based on feedback forces as in Eq.~\eqref{eq:linear_general}
reacts to a perturbation by a constant external force (Fig.~\ref{fig:det_perturb_response}e)
following a step protocol, where we consider the discretized derivative of positional data.
We observe that also in the noisy data the 
characteristic response of a system with time delay (black lines represent the deterministic response) becomes apparent 
(Fig.~\ref{fig:det_perturb_response}d).

In summary, 
inspecting Fig.~\ref{fig:det_perturb_response} and Fig.~\ref{fig:response nonlinear forces deterministic}, we notice that
all observed systems herein show a characteristic response whose details are
unique
to the specific functional form of the forces involved. 
In particular, a characteristic response is also present and visible in systems with nonlinear delayed and instantaneous forces.
We emphasize that
it is important to have a sudden onset of a sufficiently strong perturbing force 
to cause a 
sufficiently strong reaction, compared to noise in 
stochastic systems. 
Crucially, and in contrast to the PSD-based approach to inference, instead of observing the trajectory for a long time, the necessary observation duration only has to be of the order of the time delay itself.

Now the question remains, how this feature of systems with time delay can be used to infer the delay time, specifically in noisy Brownian systems. Depending on the parameters, it can be challenging to visually distinguish the characteristic response from noise, despite an average over noise realizations. 
(In Appendix \ref{sec:nonlin_perturbation_stochastic}, Fig.~\ref{fig:response nonlinear forces stochastic}, 
trajectory data for nonlinear systems are presented to give the reader some intuition concerning the complexity of responses and the obfuscation of the change in dynamical behavior through thermal noise.)
To this end, we propose the use of a neural network to classify trajectories, or rather stochastic time series, in terms of the underlying delay time. We introduce this technique in the following subsection. 
\subsection{Convolutional neural network -- time series~classification\label{sec:cnn}} 

One popular approach for classification of time series data is to employ a supervised machine learning algorithm using (deep) convolutional neural networks 
\cite{brunton_data-driven_2022,fawaz_deep_2019}.
The idea behind this approach is to train the network on labeled training data. 

A simple analogy can be found in image classification, where a network is fed labeled images of cats and dogs. Here, the goal is to correctly classify cats as cats and dogs as dogs. For this purpose, the weights and biases within the network are iteratively updated during training. The objective in training is to minimize an appropriate loss function and, at the same time, maximize classification accuracy. This simple binary example can, of course, be extended to many more, even related species (for example, distinguish between dog and wolf), making the classification more fine grained. 

Here, we use this idea to classify one-dimensional time series data (where the time series was produced by application of the perturbation method described above in Subsec.~\ref{sec:characteristic_response} and in Appendix~\ref{sec:cnn_data_prep}) in terms of the underlying delay time. We make use of the characteristic response of a system with time delay, that is, we assume that the $\tau$-dependent change of dynamical behavior within the time series is a feature the neural network can be trained on.
We focus on the discretized derivative of the positional data 
as shown in Fig.~\ref{fig:det_perturb_response}d.
In addition to the more distinct response compared to that seen in $x(t)$ (see Fig~\ref{fig:det_perturb_response}a and b),
this quantity
has the advantage that standardization and normalization (see Appendix~\ref{sec:cnn_data_prep}) is significantly better achievable compared to pure trajectory data.

\begin{figure}[h!]
    \centering
    \includegraphics[width=\columnwidth]{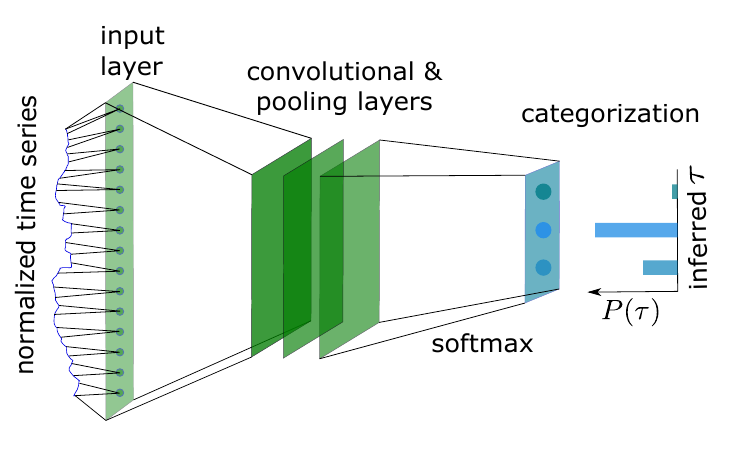}
    \caption{
    Visualization of time series classification through convolutional neural networks. Time series data (discretized derivative, see Appendix~\ref{sec:cnn_data_prep} for the input data preparation) is fed into the neural network via the input layer. 
    Then, the data goes through a sequence of convolutional and pooling layers. In our specific architecture, before the second last layer the network is flattened to be one-dimensional and subsequently fed to the last hidden layer, a densely connected layer. 
    Finally, a densely connected layer is used as a output layer, where a softmax activation function is employed, in order to return a probability distribution for an array of different $\tau_{\rm node}$. The result is a classification of the time series in terms of delay times.
}
    \label{fig:visualization_neural_network}
\end{figure}

Figure~\ref{fig:visualization_neural_network}
depicts schematically a simplified network architecture for the classification task.
It consists of an input layer which receives the data obtained from the perturbation method in terms of the discretized derivative. Several hidden convolutional and pooling layers follow. The weights and biases associated to the connections within this part of the network are modified in training to minimize an appropriate loss function.
Finally, the hidden layers are connected to the output layer, where a softmax activation function
is used such that the values of the output layer represent a discrete probability distribution $P(\tau)$ (see 
rightmost part
of Fig~\ref{fig:visualization_neural_network}) over an array of $\tau_\mathrm{node}$, i.e., the delay times used in training. In other words, each node of the output layer represents 
one 
specific
delay time $\tau_\mathrm{node}$  corresponding to one of the delay times the network was trained on. The node can assume probability values $P(\tau)$ between zero and one
with
$\sum_{\tau_{\rm node}}P(\tau)=1.$
Details on network architecture, hyperparameter tuning, and training are given in Appendix~\ref{sec:network_architecture}. 
For more information on how to train, test and predict parameters using the convolution neural networks, see 
Refs.~\cite{fawaz_deep_2019,brunton_data-driven_2022}. 

When training the network, we 
choose a range of delay times as well as a range of remaining parameters. For the linear system this implies a large number of combinations of the parameters $k_a$, $k_b$ and $\tau$, i.e., each parameter set consists of a tuple $(\tau_\mathrm{node}, k_a, k_b)$, where all are discrete values within the parameter range. The delay times are represented in the output nodes of the neural network. For each $\tau_\mathrm{node}$, the training is then carried out for many combinations of the remaining parameters. This makes the inference results robust, with the consequence that the specific combination of parameters is of minor importance for the inference accuracy of the trained network~\cite{fawaz_deep_2019}. 

Since the goal is to be able to infer time delay from noisy time series data, we train the network on several noise realizations for each of the parameter combinations. Specifically, we employ of the order of 100 realizations 
depending on the complexity of the involved delayed and instantaneous forces.

\section{Perturbation and deep learning-based approach -- numerical results\label{sec:perturb_numerical}}
In the following, we discuss results of the perturbation-based approach to infer time delay using neural networks for the trajectory data obtained from Brownian dynamics simulations (see Appendix~\ref{sec:simulation_method} for details).

We will first focus on inference by means of a convolutional neural network trained on a linear time-delayed feedback force [see Eq.~\eqref{eq:linear_general}]. With this network, we infer time delay for trajectory data obtained from numerical simulations of the linear feedback force [see Eq.~\eqref{eq:linear_general}] as a proof of concept. Subsequently, the $\tanh$ feedback force [see Eq.~\eqref{eq:tanh_fb_force}] is considered to demonstrate that the network trained on a linear feedback force is able to generalize. 
In principle, the perturbation-based approach to infer time delay
using convolutional neural networks 
trained on linear feedback forces
may also be applicable to other systems that exhibit a stationary or non-equilibrium steady state 
such as Gaussian time delayed feedback [Eq.~\eqref{eq:Gaussian_fb_force}]; 
however, an extensive analysis is beyond the scope of this study. 

Finally, we will discuss the inference results obtained from a convolutional neural network trained on trajectories based on Eqs.~\eqref{eq:dw_linear}--~\eqref{eq:tilted_washboard_nonlinear}. We use the trained network to classify trajectory data obtained from numerical simulations of the same equations of motion. In this way, we demonstrate that it is possible to train such a neural network on several functional forms of the 
(nonlinear)
delayed and instantaneous forces simultaneously while maintaining similar prediction accuracy as before. In practical applications, this allows for inference despite not knowing the exact dynamics beforehand.

\subsection{Network trained on linear model applied to linear force data\label{sec:cnn_num_lin}}
\begin{figure}[h!]
    \centering
    \includegraphics[width=\columnwidth]{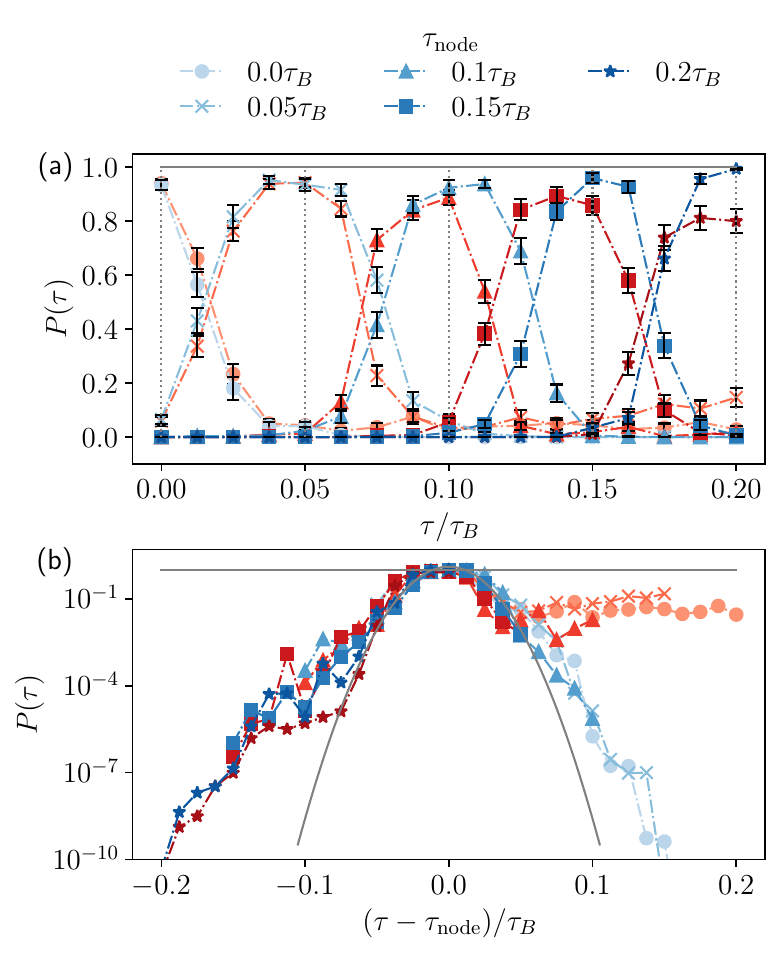}
    \caption{
    Classification probabilities for the delay time. 
    Panel (a): Colored symbols (color intensity increasing with $\tau_{\rm node}$): The probability of inferring a delay time at an exit node corresponding to $\tau_\mathrm{node}$. 
    Blue: Data corresponds to Eqs.~\eqref{eq:overdamped_Langevin} and ~\eqref{eq:linear_general} for $k_a = k_b > 0$. Red: Data corresponds to $\tanh$ feedback force~\eqref{eq:tanh_fb_force} with $A>0$. The neural network is trained on the linear model~\eqref{eq:linear_general} with parameter range $k_a\tau_\mathrm{B}^2 \in [2.5,15]$, where each input trajectory is prepared as discussed in Appendix~\ref{sec:cnn_data_prep}. We then infer time delay for both models, linear~\eqref{eq:linear_general}(blue) with same range of $k_a$ and the $\tanh$ model~\eqref{eq:tanh_fb_force}(red) with parameter range $A\tau_\mathrm{B}^2\in[4,16]$. Average over inference results for ten different trajectories (prepared as discussed in Appendix~\ref{sec:cnn_data_prep}) for each parameter combination. Each data point is an average over 
    six different values of $k_{a,b}$ and seven different values of $A$ respectively. Error bars indicate one standard error of the mean. Vertical dotted lines indicate $\tau_{\rm node}$.
    Panel (b): Same data (but log-scaled $y$-axis) as in panel (a) represented by scaling and shifting the data to reflect the deviation from the output node $\tau_{\rm node}$. Grey Gaussian curve centered around zero as well as horizontal grey line (indicating probability one) act as guides for the eye. 
    }
    \label{fig:network_linear_attractive}
\end{figure}

For our numerical study we first consider
the attractive linear feedback force system based on Eqs.~\eqref{eq:overdamped_Langevin} and 
~\eqref{eq:linear_general} with $k_a = k_b > 0$.
Here, we use a neural network that was trained on the same model to infer time delay.
In Fig.~\ref{fig:network_linear_attractive}a we show 
inference results represented by the probability distribution $P(\tau)$ (see Sec.~\ref{sec:cnn}). We averaged over the classification probabilities for different values of the parameters $k_a$ and $k_b$
at each data point, as well as over $10$ noise realizations. Each of the color-coded curves corresponds to one of the exit nodes and represents the probability of a time series to be based on the delay time value $\tau_\mathrm
{node}$. Markers correspond to the underlying delay times of test data, i.e. the time series we obtain from the perturbation approach and feed into the neural network (after applying the preparation procedure discussed in Appendix~\ref{sec:cnn_data_prep}). 
We classify time series based on delay times of $\tau = 0\tau_B$ (passive system) up to $\tau = 0.2\tau_B$. 

The highest 
classification probability 
is observed
if 
a 
time series is based on a delay time close to one 
of the $\tau_\mathrm{node}$
the network was explicitly trained on.
In those cases, we find $P(\tau)\approx 1$, that is, the network is almost certain. In close proximity to those delay times, the 
probability
remains 
high
where the resulting inference curve for each exit node approximately follows a Gaussian close to the delay values 
$\tau_\mathrm{node}$,
as shown in Fig.~\ref{fig:network_linear_attractive}b. 
For time series where $\tau$ lays inbetween two $\tau_\mathrm{node}$, the exit nodes corresponding to $\tau_\mathrm{node}$ above and below show finite probabilities.
The finer the time delay resolution in training, the better the inference results will become, up to a certain point where the network can no longer distinguish between adjacent values of $\tau_\mathrm{node}$. 
\subsection{Network trained on linear model applied to $\tanh$ feedback force data\label{sec:cnn_num_nonlin}}
For the red data points in Fig.~\ref{fig:network_linear_attractive}, we use the same network (trained on the linear feedback force with $k_a = k_b > 0$) to infer the delay time for trajectories from simulations of the $\tanh$ feedback force system, where we averaged the results for several different values of the parameter $A$ of the $\tanh$ feedback force [Eq.~\eqref{eq:tanh_fb_force}] as well as $10$ noise realizations. We ensured that the linear feedback force system was trained in a parameter range that is similar to that of the $\tanh$ feedback force, if it was linearly approximated. 

Overall, we find accurate inference results, with classification probabilities in excess of 80 percent for delay times close to those used in training. Looking closely at Fig.~\ref{fig:network_linear_attractive}a, we observe that the peaks still approximately follow a Gaussian but are shifted slightly to the left, that is, the network slightly overestimates the delay time in the test trajectories, especially for the longer delay times considered. Another observation concerns larger deviations that we find
at the highest values of the delay time $\tau$. We interpret this observation as follows. 
For long delay times, the $\tanh$ feedback force~\eqref{eq:tanh_fb_force} becomes constant at large displacements.
On the other hand, for small amplitudes $A$, the change in the feedback force in response to the perturbation
is no longer clearly distinguishable from 
thermal noise. Therefore, the underlying reason for the observed deviations may be a dominance of diffusive behavior when the $\tanh$ feedback force assumes a constant value (beyond the linear regime) for large displacements in combination with a small feedback force amplitude compared to noise.
In those cases, the neural network can no longer clearly distinguish between different delay times leading to an increased probability (see Fig.~\ref{fig:network_linear_attractive}b) for inferring short 
delay (where $\tau = 0$ corresponds to diffusive behavior). 

In summary, these results indicate that a network trained in linear time-delayed feedback is able to generalize to nonlinear systems with the underlying $\tanh$ feedback force. For suitable parameter ranges, similar results can be achieved for the Gaussian feedback force discussed above.
This indicates that systems subject to time delay that behave linearly in some limit may allow classification via a neural network trained on linear feedback forces, without any further knowledge about the functional form of the feedback forces in the systems.

\subsection{Network trained simultaneously on several nonlinear feedback forces\label{sec:full_nonlin_network}}
\begin{figure}[h!]
    \centering
    \includegraphics[width=\columnwidth]{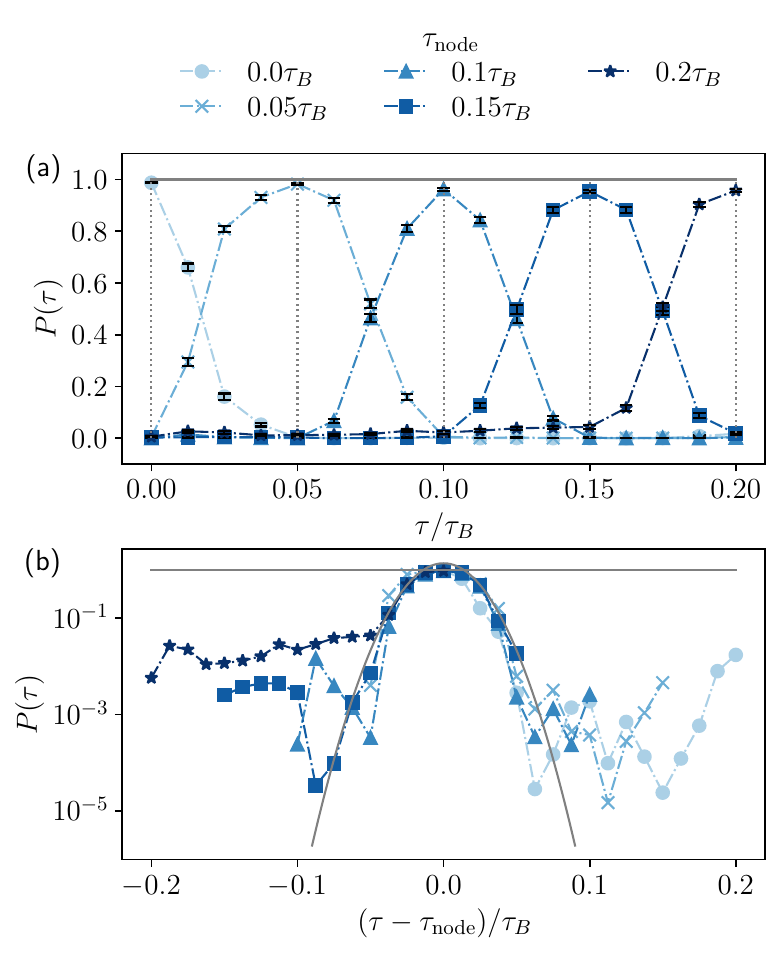}
    \caption{
    Classification probabilities for the delay time. Inference results for network trained on Eqs.~\eqref{eq:dw_linear}--~\eqref{eq:tilted_washboard_nonlinear}, where each input trajectory is prepared as discussed in Appendix~\ref{sec:cnn_data_prep}. We then infer time delay for trajectories obtained from the same model. Panel (a): Symbols (color intensity increasing with $\tau_{\rm node}$): The probability of inferring a delay time at an exit node corresponding to $\tau_\mathrm{node}$. Average over inference results for ten different trajectories (prepared as discussed in Appendix~\ref{sec:cnn_data_prep}) for each parameter combination. Each data point is an average over 
    27 
    different 
    combinations 
    of the system parameters.
    Error bars indicate one standard error of the mean. Vertical dotted lines indicate $\tau_{\rm node}$. 
    Panel (b): Same data (but log-scaled $y$-axis) as in panel (a) represented by scaling and shifting the data to reflect the deviation from the output node $\tau_{\rm node}$. Grey Gaussian curve centered around zero as well as horizontal grey line (indicating probability one) act as guides for the eye.
   }
    \label{fig:results_several_forces_network}
\end{figure}

So far, we have trained our model on only one specific feedback force in a range of parameters. However, in practical applications, it may be desirable to have a more versatile network that can infer the delay time for a wide variety of feedback forces, where the underlying functional form cannot be specified with certainty. 
To demonstrate that, in principle, this is also possible, we trained a convolutional neural network on the different nonlinear equations of motion in Eqs.~\eqref{eq:dw_linear}-\eqref{eq:tilted_washboard_nonlinear}. 
The corresponding results are presented in Fig.~\ref{fig:results_several_forces_network}. Each data point represents the classification probability averaged over $10$ 
noise realizations, 
three different functional forms of forces as well as 27 
combinations of the system parameters that are different from those used in training.
From Fig.~\ref{fig:results_several_forces_network}a we observe good inference accuracy with classification probability around 90 percent close to the delay times on which the network was trained. Furthermore, we again observe activation curves that resemble a Gaussian (see Fig.~\ref{fig:results_several_forces_network}b).
Further away from those delay times, we see an elevated activation of other output nodes, i.e. the probability of miss-classification increases between two values of $\tau_\mathrm{node}$ but overall remains small (below 10 percent).
Although it is not possible to specify the exact reason for this behavior, it might be due to the systems dynamical response in some parameter regimes being less sensitive to the perturbation applied. 

We note that while not strictly necessary, to improve inference accuracy, we here applied confining attractive harmonic instantaneous potential for a time interval longer than the delay time before applying the perturbing force to ensure a consistent initial history. 

This proof-of-concept shows that in principle the neural network can be trained on several time-delayed feedback and instantaneous forces without sacrificing inference accuracy.
Finally, in principle this method can also be used on systems with colored noise as demonstrated in Appendix~\ref{sec:colored_noise}, Fig.~\ref{fig:NN_colored_noise_inference}.

\section{Conclusions\label{sec:conclusions}}
In this article, we have proposed and demonstrated in a proof of concept two methods to infer discrete time delay in a wide range of stochastic systems with linear as well as nonlinear time-delay dependence.

Starting from linear time-delayed feedback, we have identified and analytically analyzed a characteristic signature of time delay in the PSD, which we have shown to carry over to nonlinear systems. On this basis we have developed a method to infer time delay and the specific delay time. The method requires no knowledge about the functional form of the of the underlying forces.

Since this PSD-based method requires many long stochastic trajectories for accurate inference, we asked the question (in the view of experimental situations) whether it is possible to infer time delay based on a short observation of only few realizations of a stochastic system.
To this end, we developed a perturbation-based approach that utilizes a characteristic response a system subject to time delay exhibits in the presence of an external perturbation. 
We then used convolutional neural networks to classify the resulting trajectory data in order to infer time delay for several linear and nonlinear stochastic systems subject to time delay. 
This method 
only
requires a short
duration of the observation on the order of the time delay itself. For this reason, we expect that this method, in principle, also allows inference in the absence of a non-equilibrium steady state or even in unstable systems, provided that underlying forces do not diverge during the observation time.

We emphasize that in experiments, such a perturbation can be realized in principle using state-of-the-art techniques \cite{cilibertoExperimentsStochasticThermodynamics2017a,dagoVirtualDoublewellPotential2022}. Specifically, for colloidal particles the perturbing force can be applied using optical tweezers \cite{kumarNanoscaleVirtualPotentials2018,toyabeExperimentalTestNew2007} or external fields \cite{fernandez-rodriguezFeedbackcontrolledActiveBrownian2020}.

In recent years several parameter inference techniques for dynamical systems and equation discovery (for ordinary as well as partial differential equations) from (noisy) data have been developed, methods ranging from Bayesian inference~ 
\cite{vontoussaintBayesianInferencePhysics2011}
to the application of neural network-based techniques \cite{chenPhysicsinformedLearningGoverning2021}.
However, to the best of our knowledge, only few such methods exist for systems with time delay, with highly limited applicability to stochastic systems.
Our methods to infer time delay, which can be applied to linear as well as nonlinear systems or even without explicit knowledge of the underlying forces, substantially advance the toolbox of parameter inference in stochastic systems.

In this article, we have restricted our analysis and proofs of concept to simple one-dimensional systems consisting of one Brownian particle. 
However, we 
expect
that the presented method can in principle be applied to more complex systems, which would open up many possible 
avenues for
future research and path the way to further extending the typical single colloidal particle experiments~\cite{lopezRealizationFeedbackControlled2008,bell-daviesDynamicsColloidalParticle2023,junVirtualPotentialsFeedback2012}, as %e.g., 
done in \cite{saha_cybloids_2024-1} 
where feedback with time delay is applied to several interacting colloidal particles.

Staying in one dimension, an interesting extension could be the application of the presented inference methods to partially observed systems. One possible scenario would be a passive probe particle coupled to a kinesin walker (see Appendix~\ref{sec:kinesin_walker}). Here, the question would be if, by observation and control of the probe particle alone, potential time delay in the propulsion of the kinesin walker \cite{ertelWaitingTimeDistributions2023}
could be observed.

In this study, we focused on white noise systems for our analytical and numerical considerations but also demonstrated the applicability of our inference methods to systems with exponentially correlated noise. However, in real-world systems, more complex types of noise may have to be taken into consideration (e.g., different types of colored noise, multiplicative noise, spatially inhomogeneous noise). For those systems, it remains an open challenge to
develop strategies to infer time delay by extending and generalizing the methods discussed here.

Furthermore, from a more fundamental perspective, it would be interesting to investigate whether one could exploit fluctuations-dissipation-like relations \cite{kuboStatisticalPhysicsII1991,marconiFluctuationDissipationResponse2008,gomes-filhoFluctuationDissipationRelations2025} to learn about the delay.
We note that for the present equations of motion, there is no conventional fluctuation-dissipation relation between noise correlation and memory kernel since the systems are subject to white noise (i.e., zero correlation time), whereas the `memory kernel' corresponding to discrete delay is given by a delta distribution around the delay time, $\tau$ \cite{doerriesCorrelationFunctionsNonMarkovian2021,loosMediumEntropyReduction2021}.
Still, it would be interesting to explore whether other types of fluctuation relations exist (e.g., when comparing positional moments \cite{chechkinFluctuationRelationsAnomalous2009}), 
and if they could  provide us with information about the time delay in a stochastic system.

Furthermore, 
it would be of great interest to apply the proposed methods to infer time delay in interacting colloidal many-particle systems as well as in systems from related fields, such as systems consisting of motile agents that exhibit collective behavior like flocking or swarming.
In recent years, several studies have proposed that time delay may play a role in emergent collective phenomena \cite{holubecFiniteSizeScalingEdge2021,wangSpontaneousVortexFormation2023a,pakpourDelayinducedPhaseTransitions2024}, where some have explicitly included time delay in their mathematical modeling, leading to new and interesting effects and insights. On the other hand, advances on the experimental side now make controlling many interacting colloidal particles \cite{franzlActiveParticleFeedback2020,bauerleFormationStableResponsive2020} and observing and tracking motile agents  \cite{millerSchoolingShoalingPatterns2012,partridgeSurveyingSwarmExperimental2022,kogerQuantifyingMovementBehaviour2023} possible.
Having a reliable method to infer the presence of time delay and even the delay time itself would greatly help progressing our knowledge and understanding of the mechanisms behind the emergence of collective phenomena. 

Beyond systems with one discrete delay time like those discussed here, systems with several delay times or with distributed delay are receiving considerable interest
(see Ref.~\cite{loosMediumEntropyReduction2021} 
and references therein).
Extending and generalizing the methods discussed here to such systems would be of great value and an open challenge for future research.

In summary, the 
methods presented in this article 
can be the first step towards 
an understanding of the presence and role of time delay in many experiments and real world applications, 
since knowledge about the presence of time delay is essential in modeling as well as in controlling dynamical behavior in such situations.

% \clearpage
\begin{acknowledgments}
We are grateful to Sandro Azaele (University of Padova) for many insightful and stimulating discussions. We also thank Diego Rengifo (Uniandes, TU Berlin) for carefully reading the manuscript and providing valuable feedback. This work was supported in its early stages by the Deutsche Forschungsgemeinschaft (DFG, German Research Foundation), Project No. 163436311 – SFB 910. DG gratefully acknowledges support from the Alexander von Humboldt Foundation.
\end{acknowledgments}

\appendix

\renewcommand{\thefigure}{\arabic{figure}}
\renewcommand{\appendixname}{Appendix}
\appendix
\setcounter{figure}{0}
\renewcommand{\thefigure}{\thesection\arabic{figure}} 

\section{Analytical details}
\subsection{Gaussian time-delayed feedback - analytical details}
\subsubsection{Expansion of the nonlinear feedback force in the limit of small displacements\label{sec:gaussian_linearized_small_delay}}
For small displacements, that is, small $a$ and $\tau$, the feedback force can be linearized~\cite{kopp_persistent_2023-1}, analogously to a small delay approximation~\cite{guillouzicSmallDelayApproximation1999}, resulting in a force of the form of Eq.~\eqref{eq:linear_general}, where $k_a = k_b = a/b^2$. Corresponding numerical results are presented in Sec.~\ref{sec:nonlinear_numerical_limits} of the main text.
\subsubsection{Expansion of the nonlinear feedback force in the \textit{constant velocity state}\label{sec:gaussian_linearized_const_vel}}
We can exploit the marginally stable constant velocity state \cite{kopp_persistent_2023-1} this kind of feedback force can produce in the deterministic limit. We rewrite the equation of motion in terms of position $x(t)$ relative to the position of a particle moving at constant velocity $v_\infty$. We can then linearly approximate the feedback force, assuming that the deviation is small.
This leaves us with an approximate equation of motion that describes the stochastic dynamics in the `constant velocity state' \cite{kopp_persistent_2023-1}:
\begin{equation}
	\gamma\dot{x}(t) \approx  \gamma v_\infty + K \left[x(t) - x(t-\tau) - v_\infty \tau\right] + \xi(t),\label{eq:linear_const_vel}
\end{equation}
where we define
\begin{align}
    v_\infty &= \frac{\sqrt{2}b}{\tau}\sqrt{-\ln\left(\frac{\gamma b^2}{a\tau}\right)}\ , \\
	K&\equiv \frac{a}{ b^2}\exp\left[-\frac{\left(v_\infty \tau\right)^2}{2b^2}\right]\left(1-\frac{(v_\infty \tau)^2}{b^2}\right)\ .
\end{align}
Finally, based on Eq.~\eqref{eq:linear_const_vel} we can also obtain the Fourier transform for such a system; it reads (with $K < 0$)
\begin{equation}
    \tilde x_{v_\infty}(\omega) = \frac{(\gamma + K\tau)v_\infty\delta(\omega) + \tilde \xi}{\gamma i\omega - K + K e^{-i\omega \tau}}\label{eq:lin_ana_const_vel}.
\end{equation}
In our simulations, we cannot resolve the zero-frequency contribution, which we therefore neglect in the following. Under this assumption, the Fourier transform of the position would be the same as for the linear feedback force with $k_a = k_b > 0$, which was discussed in the main text (see Sec,~\ref{sec:linear_results_analytical}).
We note that persistent motion cannot be considered a stationary state but rather a non-equilibrium steady state. The variance of the position in such systems grows over time, even if the mean particle position goes to zero for long observation times or many noise realizations. For this reason, the differencing method has to be applied in order to obtain the correct numerical results. Corresponding numerical results are presented in Sec.~\ref{sec:nonlinear_numerical_limits} of the main text.
\subsection{Definition of the power spectral density  via correlation functions\label{sec:textbook_psd}}
To put our definition of the PSD [Eq.~\eqref{eq:def_psd}] into a broader context, we note that the power spectral density is often defined via the Fourier transform of the corresponding autocorrelation function \cite{millerPowerSpectralDensity2012} under the assumption of stationarity of the underlying stochastic process. We recall that for one trajectory, the power spectral density is defined as 
\begin{equation}
    S_{xx}(f)  = \lim_{\mathcal{T}\to\infty} \frac{1}{\mathcal{T}} |\tilde{x}_\mathcal{T}(f)|^2.
\end{equation}
with frequency $f$ and observation time window $\mathcal{T}$. The quantity $\tilde{x}_\mathcal{T}$ is the Fourier transform of the position for some observation time window $\mathcal{T}$,
\begin{align}
    \tilde{x}_\mathcal{T} (f) &= \int_{-\infty}^\infty x_\mathcal{T}(t) e^{-i2\pi f t} dt \\
                    &= \int_{t_0 - \mathcal{T}/2}^{t_0 + \mathcal{T}/2} x(t) e^{-i2\pi f t} dt\ .
    \label{eq:psddef}
\end{align}
Assuming ergodicity, the right-hand side of Eq.~\eqref{eq:psddef} can be rewritten as \cite{millerPowerSpectralDensity2012}
\begin{align}
    \lim_{\mathcal{T}\to\infty} \frac{1}{\mathcal{T}}|\tilde{x}_\mathcal{T}(f)|^2 &= \int_{-\infty}^{\infty}\left[\lim_{\mathcal{T}\to\infty}\frac{1}{\mathcal{T}}\int_{-\infty}^{\infty}x_\mathcal{T}^*(t-t^\prime)
    %*
   \times  \right. \nonumber\\
    & \hphantom{=} 
    %*
    \times x_\mathcal{T}(t)dt\bigg]e^{-i2\pi ft^\prime}dt^\prime\\ 
    &= \int_{-\infty}^{\infty}R_{xx}(t^\prime)e^{-i2\pi ft^\prime}dt^\prime \ ,
\end{align}
with $R_{xx}$ being the autocorrelation function of the position $x(t)$.
For, e.g., a confined Brownian particle, the mean and variance of the position are constant after some time. Under this condition, the power spectral density can be computed via the Fourier transform of the corresponding positional autocorrelation function as well as via the modulus of the Fourier transform of the position. 
We fully rely on the latter since the autocorrelation is only of limited use in analytical computations for the time delay systems considered here. Furthermore, the Fourier transform of the position of a linear system can be easily computed both analytically and numerically. 
\subsection{Linear time-delayed feedback - small frequency behavior \label{sec:small_freq}}
We consider the limit of $\omega \ll 1/\tau$ for $k_a\neq k_b$, and expanding to second order in $\omega\tau$ this gives
\begin{align}
    \langle S(\omega)\rangle_\mathrm{lin} \approx \dfrac{2 \gamma k_{\rm B}T}{ \omega^2k_a k_b(\tau - \tau_+)(\tau-\tau_-) + (k_a - k_b)^2}\ ,
    \label{eq:approx_psd}
\end{align}
where we defined 
\begin{align}
    \tau_\pm \equiv  \gamma\frac{1\pm \sqrt{1-k_a/k_b}}{k_a}\ .
\end{align}
From the approximation (Eq.~\ref{eq:approx_psd}) we can see that the $\mathrm{PSD}$ scales as $1/\omega^2$. 
For $k_{a,b}>0$, this behavior can be interpreted as a reduced ability of the feedback trap to confine compared to the instantaneous harmonic trap, resulting in a corresponding deviation from the black curve in Fig.~\ref{fig:PSD_analytical_var_kb} at small $\omega$ due to diffusive behavior in the long time limit. 
Analogously, for $k_{a,b} < 0$ the frequency dependency is connected to enhanced diffusion due to repulsive time-delayed feedback

\subsection{Linear time-delayed feedback - marginally stable system\label{sec:marginally_stable_app}}
The marginally stable case where $k_a=k_b$, is special and has been the subject of several previous studies \cite{kopp_persistent_2023-1,saha_cybloids_2024-1}. 
Its peculiar properties are reflected by the PSD (solid blue and dashed green lines in Fig.~\ref{fig:PSD_analytical_var_kb}). 
For the attractive feedback, in the limit of $k_a = k_b > 0$, the system essentially shows diffusive behavior for all frequencies ($\mathrm{PSD} \propto 1/\omega^2$), indicating that in this case the ability of the feedback to trap the particle has been lost.
For repulsive feedback, on the other hand, i.e., for $k_{a,b} < 0$ when $k_a=k_b$, we observe an enhanced diffusion (higher value of the PSD at small frequencies), in agreement with previous studies on this type of system \cite{kopp_persistent_2023-1,saha_cybloids_2024-1}. 
Otherwise, the repulsive feedback can have a confining effect, albeit less strict than a harmonic trap, as can be seen for $|k_a| < |k_b|$ at small frequencies (cyan and grey dashed lines in Fig.~\ref{fig:PSD_analytical_var_kb}), where the PSD resembles that of a harmonically trapped particle.

\section{Numerical procedure}
We perform numerical simulations, data analysis, and data visualization using the Python programming language. For numerical computations and data processing, we use \textit{numpy} \cite{charlesr.harrisetal.ArrayProgrammingNumPy2020} and \textit{scipy} \cite{paulivirtanenetal.SciPy10Fundamental2020}, and for visualization, we use \textit{matplotlib} \cite{Hunter:2007}. The deep learning computations are implemented in \textit{keras} \cite{francoischolletetal.Keras2015}, based on \textit{tensorflow} \cite{martinabadietal.TENSORFLOWLargeScaleMachine2015}.
\subsection{Numerical solution of the overdamped delay Langevin equation\label{sec:simulation_method}}
To solve the 
SDDE~\eqref{eq:overdamped_Langevin}, we perform BD simulations based on the Euler-Maruyama integration scheme.
\cite{maruyamaContinuousMarkovProcesses1955,kloedenNumericalSolutionStochastic1992}. 
The resulting discretized version of Eq.~\eqref{eq:overdamped_Langevin} reads 
\begin{eqnarray}
    x_{n+1} = x_{n} + \gamma^{-1}F(x_{n}, x_{n-N_\tau})\Delta t + \nonumber\\
    + \gamma^{-1}\sqrt{2k_B T \gamma \Delta t}~\eta_{n}\ , \label{eqn:sim}
\end{eqnarray}
where $N_\tau = \tau/\Delta t$ is the number of time steps within one delay time $\tau$, for the discretization time step $\Delta t$. 
$\eta_n$ is a uncorrelated Gaussian random number with zero mean and unit variance, i.e., $\langle \eta_n\eta_m\rangle = 0$ for $n\neq m$.  
Equation~\eqref{eqn:sim} is solved provided an initial history for the time interval $[-\tau, 0]$ via the discretized history function $\Phi(t_n) = \Phi_{t_n}$, that is, the trajectory in the discretized time interval $[-\tau, 0]$, has to be known.
For all stochastic simulations in this study, we used the trajectories of free Brownian particles as the initial history. 
Furthermore, we used time steps of $\Delta t = 10^{-4}\tau_B$ (much smaller than the smallest delay time considered) 
to ensure sufficient numerical precision in all the cases considered.
\subsection{The Differencing method\label{sec:differencing}}
For the computation of the (noise-averaged) PSD~\eqref{eq:naPSD_definition} a stationary time series is required, which is quite often not the case in physical systems. Already for a passive Brownian particle this problem has to be addressed, for feedback-driven systems it can become even more pronounced.

\begin{figure}[h]
    \centering
    \includegraphics[width=\columnwidth]{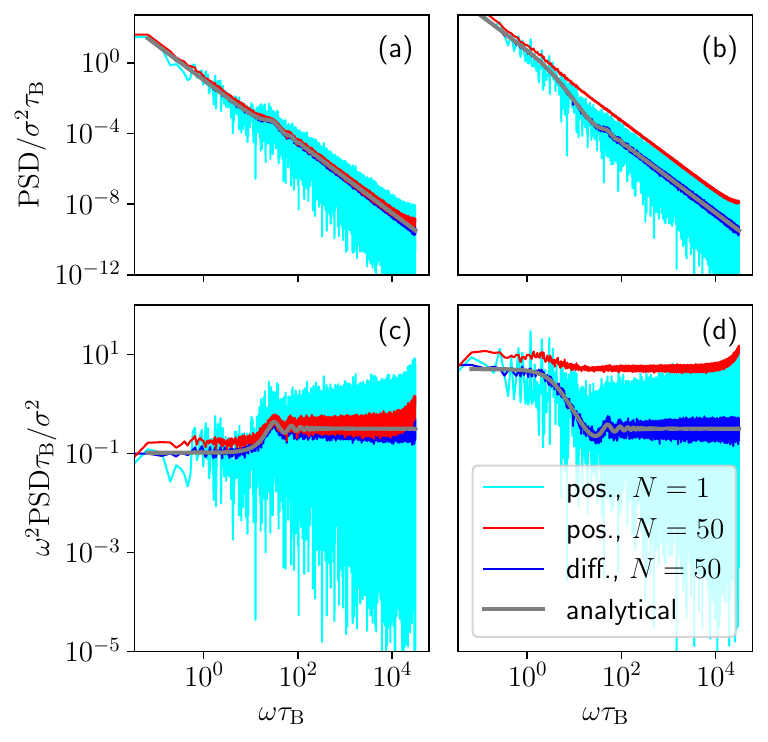}
    \caption{Comparison of position based and differencing based computation of $\mathrm{PSD}$ and $\omega^2\mathrm{PSD}$ for a system with linear feedback force [Eq.~\eqref{eq:linear_general}]. Panels (a) and (c): Marginally stable system with attractive harmonic feedback; panels (b) and (d) are for repulsive harmonic feedback below the deterministic threshold, i.e., $|k_a| = |k_b|< \gamma/\tau$.}
    \label{fig:differencing_method_numerical}
\end{figure}

Figure~\ref{fig:differencing_method_numerical} 
presents exemplary data for systems with attractive and repulsive harmonic time-delayed feedback which both behave diffusively, especially in the long time limit. The cyan and red 
data correspond to
PSD 
computed 
by directly applying the Fourier transform to the nonstationary positional data.
We observe a shift compared to the analytical expectation that is hardly visible in the PSD but becomes apparent in $\omega^2$PSD. While when averaging over a small number of noise realizations, this shift is lost in noise, when averaging over a larger number of noise realizations, the shift becomes obvious. 

The reason behind this behavior can be understood when considering that we are taking the Fourier transform of a (within the observation period) not strictly ergodic and certainly non-stationary quantity. 
More specifically, the time series is non-stationary since the variance of the position grows over time. 
Changing the temporal resolution or the length of the trajectory only has a small effect on the observed shift. 
The observed shift can intuitively be understood as follows: In a stationary system (assuming zero mean and constant variance), the position will (more or less) fluctuate around zero, given that the observation window is long enough.
However, if the process is not stationary in the sense that the variance grows as a function of time, in some realizations, the position will be positive (or negative) for a significant part of the observation time. When taking the Fourier transform of such a trajectory, it will be somewhat biased. If we took the average over noise realizations now, we could expect this bias to average out over different realizations. However, following the definition of the PSD~\eqref{eq:naPSD_definition} we compute the square modulus of the Fourier transform of the position before applying the noise average. For a stationary process, this does not matter. However, in our case, the bias can not be averaged out but rather accumulates and depends on the variance of the process and how much of the observation period will be uninterruptedly spent on one side of the mean position. 
We therefore have to approach the problem differently and make the time series stationary before even computing the Fourier transform in order to obtain results that match our analytical expectation.

To this end,
we use the differencing method, that is, instead of using the positions as time series, we consider the positional difference over a fixed time interval $\Delta t$. This is a textbook technique from time series analysis \cite{shumwayTimeSeriesAnalysis2017} to make non-stationary time series stationary. For our case, instead of computing the Fourier transform of the position we rather compute the Fourier transform of the positional difference, i.e., 
\begin{equation}
    \dfrac{\widetilde{\Delta x(t)}}{\Delta t} = \dfrac{\mathcal{F}\big[x(t)-x(t-\Delta t)\big]}{\Delta t}\ ,
\end{equation}
where the operator $\mathcal{F}$ represents the Fourier transform.
Intuitively, this looks like the Fourier transform over a temporal derivative (which we already exploited in the analytical considerations) or essentially (for small $\Delta t$) the Fourier transform of the right-hand side of the discretized Langevin equation.
Applying this method, we can obtain a numerical result that matches exactly the analytical expectation, as shown by the dark blue curves in Fig.~\ref{fig:differencing_method_numerical}.

The differencing method was used for all results discussed in Sec.~\ref{sec:psd_results} of the main text.

\subsection{Extraction of frequencies corresponding to minima and maxima of the PSD\label{sec:psd_procedure}}
In the following, we describe a procedure that allows us to extract the delay time from simulation data, which was used in Sec.~\ref{sec:full_nonlin} of the main text. It is based on the analytical results obtained for the linear system, but can also be applied to systems that depend on $x(t)$ and/or $x(t-\tau)$ in a nonlinear fashion. 
This procedure relies on the spacing of the positions of maxima and minima in the quantity $\omega^2\mathrm{PSD}$, and is applicable as long as the signal-to-noise ratio in the system allows for reliable determination and extraction of the position of these minima and maxima. As such, the larger the number of available noise realizations, the better the inference results. %In
For example, Fig.~\ref{fig:differencing_method_numerical} illustrates the level of noise that has to be dealt with in this kind of system in the case of a marginally stable time delay system. In order to obtain a good polynomial fit over multiple peaks, we need to average over a sufficient number of noise realizations; for the figures in the main text [Figs.~\ref{fig:interpolation}a-c] 400 noise realizations were used. 

While, in principle, the positions of the minima can be obtained manually, it is more reliable and convenient to use a spline interpolation for a subsequent semi-automatic extraction of the positions of minima and maxima. It is semi-automatic in the sense that one has to determine for how many minima and maxima a reliable fit can be obtained. 
Specifically, we use quartic splines for fitting the numerical data, which allows us to extract the roots of its derivative automatically. As the data are noisy and somewhat irregular, special care has to be taken when adjusting the smoothing parameter in the spline interpolation, such that the interpolation faithfully represents the location of the minima and maxima in the numerical data. Furthermore, limiting the range of the interpolation can significantly improve the results since at low frequencies additional oscillations with different oscillation period are observed in some cases and at high frequencies the oscillations are lost in noise. 

Specifically, the extraction of the delay time requires the following steps:
\begin{enumerate}
    \item Apply spline interpolation to $\omega^2\mathrm{PSD}$ using quartic splines implemented within \textit{scipy}.
    \item Take derivative of the resulting polynomial (built-in function of spline interpolation in \textit{scipy}).
    \item Compute the roots $\omega_n$ of the derivative of the polynomial corresponding to the minima and maxima in $\omega^2 {\rm PSD}$ (built-in function of spline interpolation in \textit{scipy}).
    \item Discard the first $N_\mathrm{discard}$ minima and maxima (until the minima and maxima have approximately equal spacing) (see Sec.~\ref{sec:linear_results_analytical}).
    \item Determine $N_\mathrm{ext}$, the number of maxima and minima that are reliably represented by the polynomial; terminate the fit before the oscillations are lost in noise.
\end{enumerate}
The inferred delay time is then given by 
\begin{equation}
    2\pi/\tau_\mathrm{inferred} =  \frac{\sum_{n=1}^{N_\mathrm{ext}-2} \left(\omega_{n+2}-\omega_{n}\right)}{(N_\mathrm{ext}-2)}\ ,
\end{equation}
where $\tau_\mathrm{inferred}$ is the inferred delay time, $N_\mathrm{ext}$ is the number of minima and maxima that can be reliably identified, and we averaged the distances
between all available pairs of consecutive minima and maxima.

\subsection{Deep learning: data preparation\label{sec:cnn_data_prep}}
For the perturbation-based deep learning approach, we first need to prepare the data. We apply the following steps:
\begin{enumerate}
    \item Take an ensemble average over 25 noise realizations over the positional data 
    at each point in time.
    \item Apply the differencing method, $\{\bar{x}(t)\} \equiv \{x(t+dt)\}-\{x(t)\}$, to render the time series stationary.
    \item Apply a simple moving average over three subsequent time steps for denoising.
    \item Normalize by dividing by the maximum value during the observation period.
\end{enumerate}
The data are prepared in this way for training the neural network as well as for inference using the trained network (see Sec.~\ref{sec:perturb_numerical}).

\subsection{Deep learning: network architecture and hyperparameter tuning\label{sec:network_architecture}}
\begin{figure}
    \centering
    \includegraphics[width=\columnwidth]{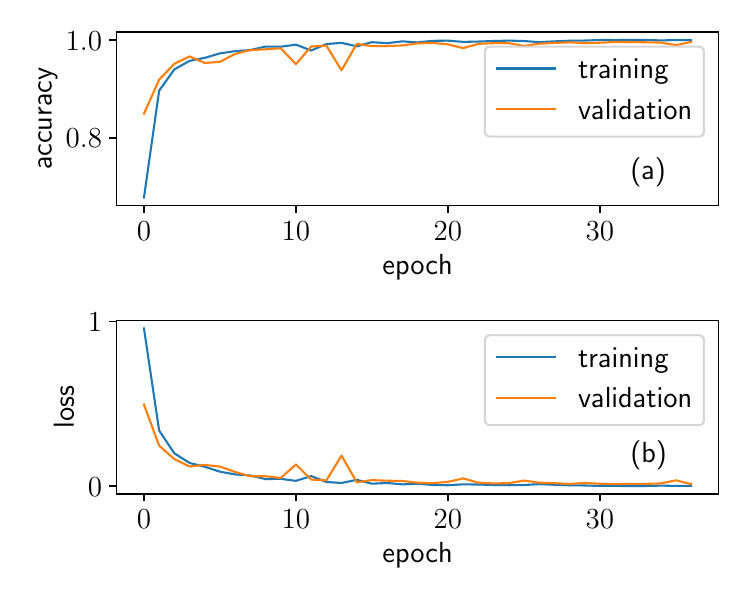}
    \caption{
    Training the convolutional neural network on nonlinear feedback forces [Eqs.~\eqref{eq:dw_linear}–\eqref{eq:tilted_washboard_nonlinear}]: Panel (a) shows the classification accuracy, and panel (b) shows the training loss (computed via sparse categorical crossentropy), both plotted as functions of training epochs for the training and validation datasets.}
    \label{fig:training_acc_loss}
\end{figure}

For the classification of trajectory data we use a convolutional deep neural network (see e.g., \cite{brunton_data-driven_2022,fawaz_deep_2019}) implemented using \textit{keras} \cite{francoischolletetal.Keras2015} based on \textit{tensorflow} \cite{martinabadietal.TENSORFLOWLargeScaleMachine2015}. The neural network (see Sec.~\ref{sec:cnn}, especially  Fig.~\ref{fig:visualization_neural_network} for a simplified schematic depiction) consists of an input layer (the number of nodes equals the number of relevant time steps), hidden layers, and an output layer (the number of nodes equals the number of classes, i.e., delay times the network is trained on). 
For hidden layers, considerations such as complexity, regularization, and available resources play a crucial role. Here, we focus on a three convolutional layers and pooling layers to limit complexity. We use a hyperparameter tuning scheme in order to achieve the desired accuracy for classification by the network.
This means that over how many nodes pooling is applied in a pooling layer, filter and kernel size in convolutional layers, the optimization method and the activation function are chosen automatically.
Before the final hidden layer, the network is flattened. The final hidden layer is then a densely connected layer followed by the densely connected output layer. The number of nodes of the output layer is equal to the number of different delay times on which the model is trained. A \textit{softmax} activation function is used to obtain a classification in terms of the probability that the classified trajectory is based on a certain delay time.
The network is then constructed as shown in Table~\ref{tab:network_architecture}.
\begin{table}[]
    \centering
    \begin{tabular}{|c||c|c|c|}
        \hline
         layer & role/structure & activation \\\hline\hline
         1 & input & -  \\
         2 & normalization & - \\
         3 & convolutional & relu \\
         4 & max pooling & - \\
         5 & convolutional & relu \\
         6 & max pooling & - \\
         7 & convolutional & relu \\
         8 & max pooling & - \\
         9 & flatten & - \\
         10 & dense & relu \\
         11 & output/dense & softmax\\
         \hline
    \end{tabular}
    \caption{Network structure by layers; hyper-parameters, i.e., over how many nodes pooling is applied in a pooling layer, filter and kernel size in convolutional layers, the optimization method and the activation function are optimized by hyperparameter tuning. For the input data preparation, see Appendix~\ref{sec:cnn_data_prep}. 
    }
    \label{tab:network_architecture}
\end{table}
As the loss function, which is minimized during training, the sparse categorical cross-entropy was used.
Figure~\ref{fig:training_acc_loss} presents exemplary accuracy [Fig.~\ref{fig:training_acc_loss}(a)] and loss [Fig.~\ref{fig:training_acc_loss}(b)] as a function of training epochs for the training of the neural network on several nonlinear feedback forces [Eqs.~\eqref{eq:dw_linear}-\eqref{eq:tilted_washboard_nonlinear}] simultaneously. We observe good convergence already at as little as 20 training epochs.

\subsection{Characteristic response of a nonlinear feedback force system\label{sec:nonlin_perturbation}}
\begin{figure}
    \centering
    \includegraphics[width=\columnwidth]{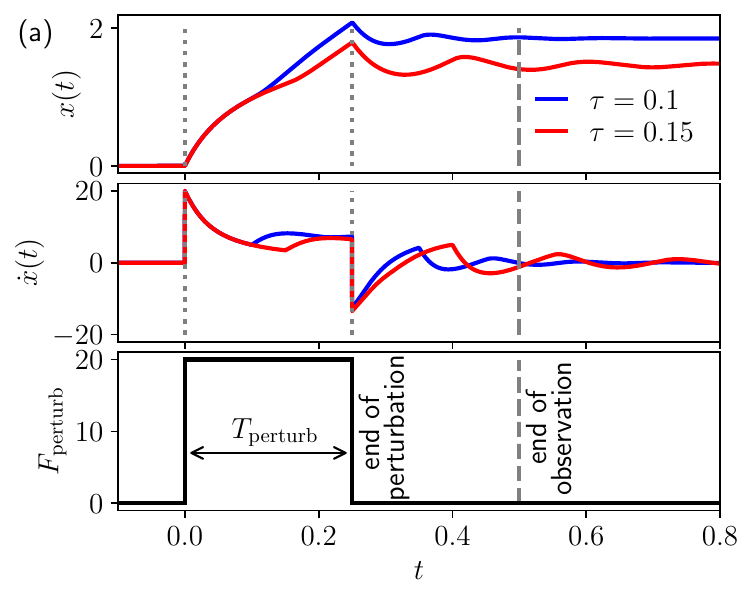}
    \includegraphics[width=\columnwidth]{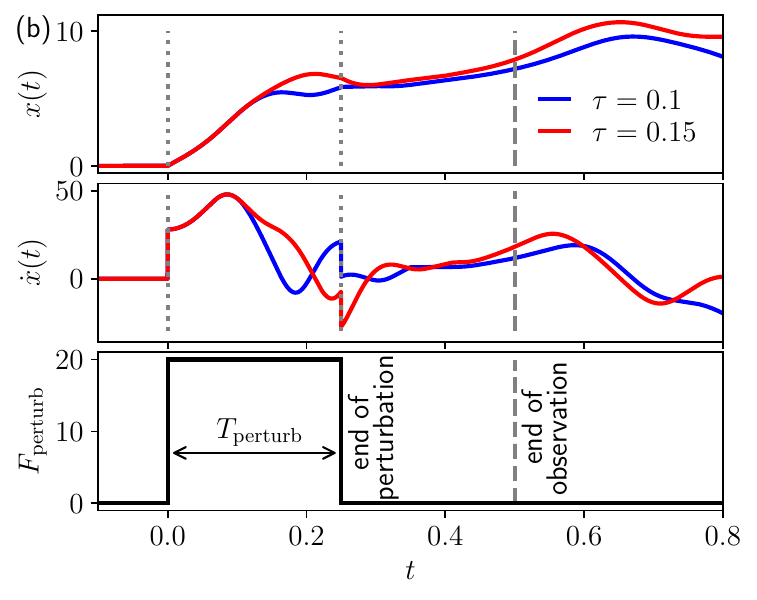}
    \caption{
    Characteristic response of a nonlinear system with time delay to an external perturbation (see also Sec.\ref{sec:characteristic_response}). Panel (a): Response of the $\tanh$ feedback force [Eq.\eqref{eq:tanh_fb_force}] to a constant external force. Panel (b): The response of a system based on Eq.~\eqref{eq:tilted_washboard_nonlinear} to the same type of perturbation. Vertical dotted lines indicate the onset and termination of the perturbation, while the vertical dashed line marks the end of the observation window.}
    \label{fig:response nonlinear forces deterministic}
\end{figure}

Figure~\ref{fig:response nonlinear forces deterministic} presents exemplary data for the response of systems with nonlinear feedback forces. In all cases, a complex but still characteristic response is observed. As for linear systems, a change in dynamical behavior is observed one delay time after switching on the perturbation as well as one delay time after switching off the perturbation.
\subsection{Characteristic response of a stochastic system with nonlinear feedback force 
\label{sec:nonlin_perturbation_stochastic}}
\begin{figure}
    \centering
    \includegraphics[width=\linewidth]{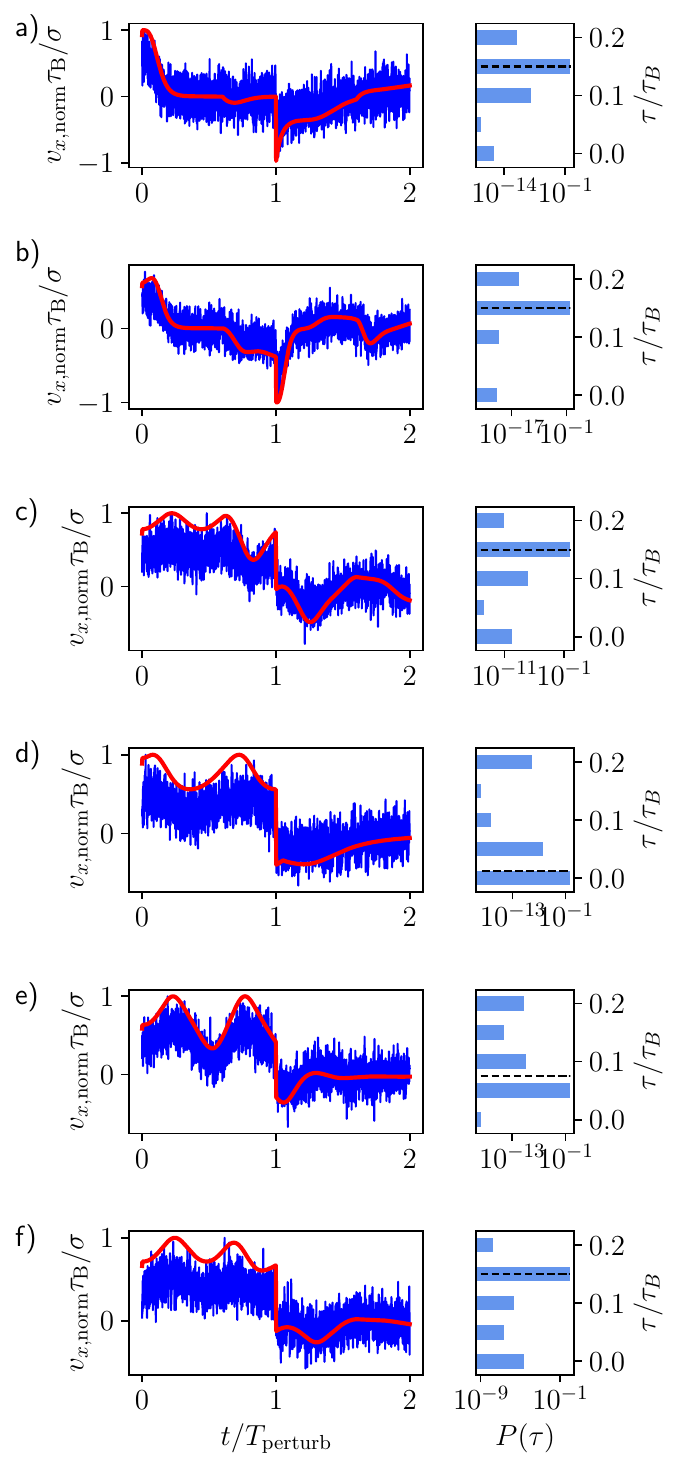}
    \caption{
    Characteristic response of a stochastic nonlinear system with time delay to an external perturbation (see also Sec.~\ref{sec:characteristic_response}) and corresponding inference results (right panels). Trajectories prepared as discussed in Appendix~\ref{sec:cnn_data_prep}. Solid line: corresponding deterministic response. Dashed horizontal line: true delay time.  Panels a)-c): $\tau = 0.15\tau_B$; a) Eq.~\eqref{eq:dw_linear} at $a = 13k_\mathrm{B}T/\sigma^4$, $b = 3k_\mathrm{B}T/\sigma^2$, $c = 11k_\mathrm{B}T/\sigma^2$; b) Eq.~\eqref{eq:dw_nonlinear} at $a = 13k_\mathrm{B}T/\sigma^4$, $b = 18k_\mathrm{B}T/\sigma^2$, $c = 17k_\mathrm{B}T/\sigma^3$; c) Eq.~\eqref{eq:tilted_washboard_nonlinear} at $a = 7k_\mathrm{B}T/\sigma$, $b = 10k_\mathrm{B}T/\sigma$, $c = 17k_\mathrm{B}T/\sigma$. Panels d)-f): Trajectories based on Eq.~\eqref{eq:tilted_washboard_nonlinear} at d) $\tau = 0.0125\tau_B$, $a = 7k_\mathrm{B}T/\sigma$, $b = 10k_\mathrm{B}T/\sigma$, $c = 17k_\mathrm{B}T/\sigma$; e) $\tau = 0.075\tau_B$, $a = 13k_\mathrm{B}T/\sigma$, $b = 3k_\mathrm{B}T/\sigma$, $c= 11k_\mathrm{B}T/\sigma $; f) $\tau = 0.15\tau_B$, $a = 9k_\mathrm{B}T/\sigma$, $b = 3k_\mathrm{B}T/\sigma$, $c = 7k_\mathrm{B}T/\sigma$. 
    }
    \label{fig:response nonlinear forces stochastic}
\end{figure}

Figure~\ref{fig:response nonlinear forces stochastic} presents exemplary data for the response of stochastic systems~\eqref{eq:dw_linear}-\eqref{eq:tilted_washboard_nonlinear} subject to nonlinear and delayed forces with corresponding inference results. In contrast to the linear system (for comparison see Fig.~\ref{fig:det_perturb_response}), the response of a nonlinear system can be more complex, 
as indicated by the solid lines representing the deterministic response in Fig.~\ref{fig:response nonlinear forces stochastic}. In the stochastic data, the changes in dynamical behavior that constitutes the characteristic response is obfuscated by thermal noise. Still, the neural network can infer time delay with high accuracy for delay times the network was trained on for trajectories prepared as discussed in Appendix~\ref{sec:cnn_data_prep}. 
For delay times in-between two subsequent $\tau_\mathrm{node}$, the inference results tend to be less accurate. 
\section{Extension to a system with harmonically coupled probe particle\label{sec:kinesin_walker}}
Here we consider a one-dimensional system of a Brownian particle subject to time-delayed feedback, which is coupled to a passive probe particle via an attractive harmonic potential.
We refer to the feedback-driven particle using the superscript $\mathrm{fb}$ and to the probe particle using the superscript $\mathrm{pr}$.
The dynamics of the feedback-driven particle is given by an overdamped Langevin equation. The tracer particle is coupled to the feedback driven particle via the position $x_\mathrm{fb}$, which enters the harmonic coupling potential.
The coupled equations of motion then read
\begin{align}
    \gamma_\mathrm{fb} \frac{dx_\mathrm{fb}}{dt} &= F_\mathrm{fb}\left(x_\mathrm{fb}(t),x_\mathrm{fb}(t-\tau)\right) - F_{\mathrm{c}} +\xi_\mathrm{fb}(t),\label{eq:coupled_1} \\
    \gamma_\mathrm{pr} \frac{dx_\mathrm{pr}}{dt} &= F_\mathrm{c} +\xi_\mathrm{pr}(t),
    \label{eq:coupled_2}
\end{align}
with $F_{\mathrm{c}} = F_\mathrm{c}\left(x_\mathrm{pr}(t),x_\mathrm{fb}(t)\right)$.
The constants $\gamma_\mathrm{fb}$ and $\gamma_\mathrm{pr}$ are the respective friction constants with $\gamma_\mathrm{fb} = \gamma_\mathrm{pr} = \gamma$ for simplicity. 
The coupling potential $U$ reads
\begin{equation}
    U(x_\mathrm{pr}(t),x_{fb}(t)) = \frac{k}{2} \left(x_\mathrm{pr}(t) - x_\mathrm{fb}(t)\right)^2,
\end{equation}
resulting in the force $F_\mathrm{c} = -\nabla_{x_\mathrm{pr}} U$ and analogously $-\nabla_{x_\mathrm{fb}} U = -F_\mathrm{c}$ which acts on the coupled probe particle.
Furthermore, $\xi_\mathrm{fb}$ and $\xi_\mathrm{pr}$ represent a one-dimensional Gaussian white noise, respectively, with zero mean and correlation function $\langle \xi_{\mathrm{fb}/\mathrm{pr}} (t) \xi_{\mathrm{fb}/\mathrm{pr}}(t^\prime) \rangle = 2 \gamma_{\mathrm{fb}/\mathrm{pr}} k_BT \delta(t-t^\prime)$.
Then, the Fourier transform of the dynamics reads as follows.
\begin{align}
    \gamma i\omega \tilde{x}_\mathrm{fb}(\omega) &= &-[k_a\tilde{x}_\mathrm{fb}(\omega) - k_b e^{-i\omega \tau}\tilde{x}_\mathrm{fb}(\omega)] \nonumber\\
    &&- k [\tilde{x}_\mathrm{fb}(\omega) - \tilde{x}_\mathrm{pr}(\omega)] + \tilde\xi_\mathrm{fb}(\omega), \\
    \gamma i\omega \tilde{x}_\mathrm{pr}(\omega) &= &k [\tilde{x}_\mathrm{fb}(\omega) - \tilde{x}_\mathrm{pr}(\omega)] + \tilde\xi_\mathrm{pr}(\omega).
\end{align}
We can now solve for $\tilde{x}_\mathrm{fb}$ and $\tilde{x}_\mathrm{pr}$ respectively and obtain
\begin{align}
    \tilde{x}_\mathrm{fb} &= \frac{\frac{k\tilde{\xi}_\mathrm{pr}}{\gamma i \omega + k} + \tilde \xi_\mathrm{fb}}{\gamma i\omega + k_a - k_b e^{-i\omega \tau} +k   - \frac{k^2}{\gamma i \omega + k}}, \label{eq:coupled_fb}\\
    \tilde{x}_\mathrm{pr} &= \frac{(k \tilde{x}_\mathrm{fb} + \tilde{\xi}_\mathrm{pr})}{\gamma i\omega + k}. \label{eq:coupled_probe}
\end{align}
The analytical expression for the PSD then follows from the definition as for the simple linear system.
From Eqs.~\eqref{eq:coupled_fb} and \eqref{eq:coupled_probe} we can already see 
that the characteristic oscillations, and, thus, the signature of time delay extend to the probe particle. Therefore and in principle, by observing the probe particle alone, inference of time delay is possible with a method similar to that proposed in Sec.~\ref{sec:perturb_method} of the main text.
\section{Systems with exponentially correlated noise -- numerical results\label{sec:colored_noise}}
In this section, we demonstrate that our methods of inferring the time delay are not limited to white noise systems but can be applied or extended to systems with colored noise. 
In other words, 
the signatures of time delay remain 
intact
even in colored-noise systems.

To this end, we first introduce exponentially correlated colored noise and subsequently present exemplary data for the (power spectral density) PSD-based inference approach as well as the neural network-based approach for non-linear systems. 
Specifically, 
we consider the
nonlinear systems in 
in Eqs.~\eqref{eq:dw_linear}-\eqref{eq:tilted_washboard_nonlinear}
and replace the Gaussian white noise (present on their right-hand side) by an additive colored noise~\cite{hanggiColoredNoiseDynamical1994}
\begin{equation}
    \dot{\zeta}(t) = -\frac{1}{t_\mathrm{corr}} \zeta(t) + \sqrt{\frac{2 \gamma k_\mathrm{B}T}{t_\mathrm{corr}^2}}~\eta(t)\ ,
\end{equation}
where $t_\mathrm{corr}$ is the correlation time and $\eta(t)$ is a white noise (delta-correlated) with zero mean and unit variance.

It turns out that in the long-time limit (i.e., the stationary state), the colored noise $\zeta(t)$ is exponentially correlated:
\begin{align}
    \langle \zeta(t)\zeta(t')\rangle_{\rm ss} = \frac{\gamma k_\mathrm{B}  T}{t_\mathrm{corr}} e^{-|t-t'|/t_{\rm corr}}\ . \label{col-cov}
\end{align} 
Before proceeding further, we would like to highlight its two limiting behaviors. 1) In the limit $t_\mathrm{corr}\to 0$, for $t =t'$, the %following 
covariance~\eqref{col-cov} diverges; otherwise, the 
correlation goes to $0$ for $t\neq t'$. Further, 
integrating the right-hand side of Eq.~\eqref{col-cov} from $t\in (-\infty, +\infty)$ gives $2\gamma k_{\rm B}T$. %So
Therefore, this correlation is a delta function in this particular limit (with a prefactor $2\gamma k_\mathrm{B}T$). 2) In the limit $t_\mathrm{corr}\to \infty$, $\dot {\zeta}\to 0$, i.e., $\zeta$ 
becomes 
a quenched random Gaussian variable with zero mean and variance $\gamma k_\mathrm{B}T/(t_{\rm corr})\to 0$ (as $t_{\rm corr}\to \infty)$.

\begin{figure}
    \centering
    \includegraphics[width=\linewidth]{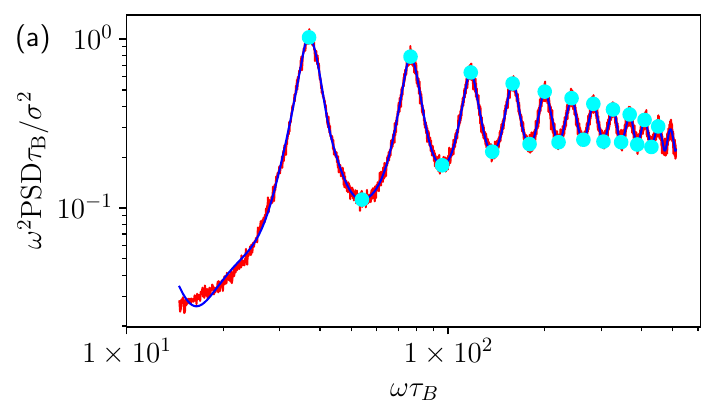}
    \includegraphics[width=\linewidth]{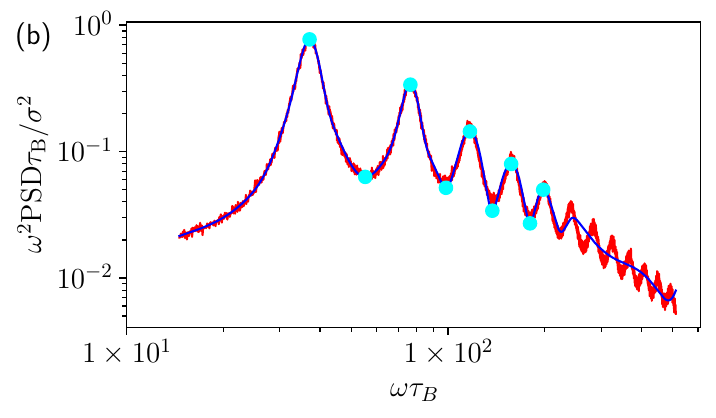}\\
    \includegraphics[width=\linewidth]{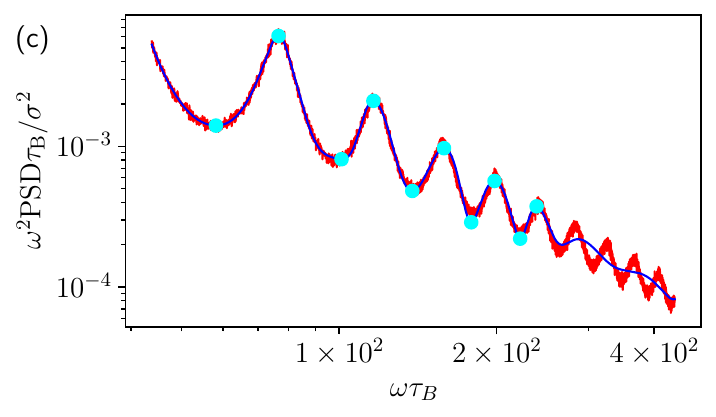}\\
    \includegraphics[width=\linewidth]{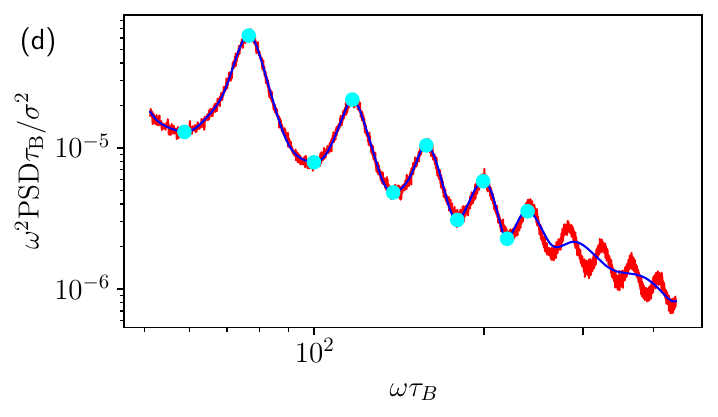}
    \caption{
    Numerically simulated PSD (red) as a function of $\omega$ for non-linear systems~\eqref{eq:dw_nonlinear} with white noise replaced by colored noise.  
    In each panel: Polynomial fit (blue) using the quartic splines and automatically extracted positions (cyan circles) of maxima and minima using the roots of the derivative of the polynomial (see Appendix \ref{sec:psd_procedure}). See Table~\ref{tab:inferred_delay_col_noise_fig} for the corresponding inference results.
    Colored noise at correlation times: a) $t_\mathrm{corr}=0.01 \tau$, b) $t_\mathrm{corr}=0.1 \tau$, c) $t_\mathrm{corr}=1 \tau$ and d) $t_\mathrm{corr}=10 \tau$, other parameters as in Fig.~\ref{fig:interpolation} and Table~\ref{tab:inferred_delay_times}.}
    \label{fig:PSD_colored_noise_inference}
\end{figure}

Figure~\ref{fig:PSD_colored_noise_inference} 
shows the PSD and the extracted positions of minima and maxima. 
Table~\ref{tab:inferred_delay_col_noise_fig} 
presents the corresponding inference results for one delay time at different values of the correlation time $t_\mathrm{corr}$.
Interestingly, while the behavior of the PSD as a function of the frequency changes as the colored noise is 
added to the system and the correlation time $t_\mathrm{corr}$ is increased, the characteristic signature of time delay, i.e., oscillations with a delay-specific oscillation period, remains.

\begin{table}[h!]
    \centering
    \begin{tabular}{|c|c|c|c|}
        \hline 
        $t_\mathrm{corr}/\tau$ & $\tau/\tau_{\rm B}$ &  $\tau_\mathrm{inferred}/\tau_{\rm B}$ & $\sigma_\tau/\tau_{\rm B}$ \\
        \hline 
        $0.01$     &0.15      &  $0.151$             & $\pm 0.003$ \\
        $0.1$     &0.15      &  $0.152$             & $\pm 0.007$ \\
        $1$    &0.15           &  $0.153$             & $\pm 0.009$ \\
        $10$      &0.15       &  $0.152$             & $\pm 0.010$ \\
        \hline
    \end{tabular}
    \caption{
    Inference of the delay times corresponding to system in Eq.~\eqref{eq:dw_nonlinear} 
    based on the PSD-data shown in~Fig.~\ref{fig:PSD_colored_noise_inference}.
    }
    \label{tab:inferred_delay_col_noise_fig}
\end{table}

\begin{figure*}
    \centering
    \includegraphics[width=.49\linewidth]{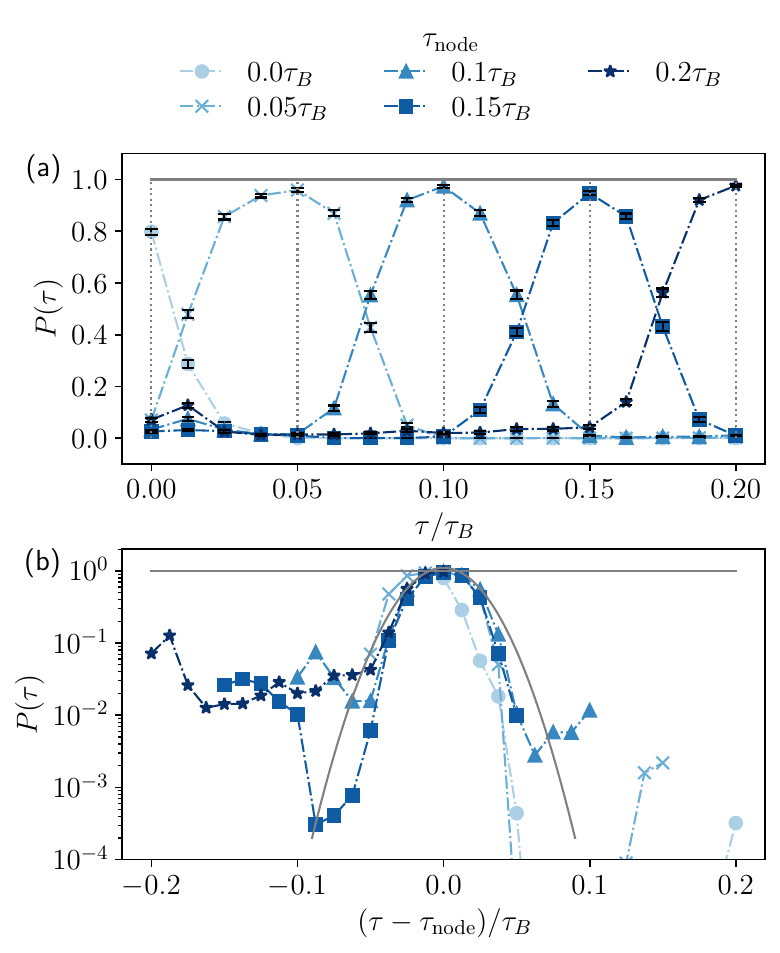}~
    \includegraphics[width=.49\linewidth]{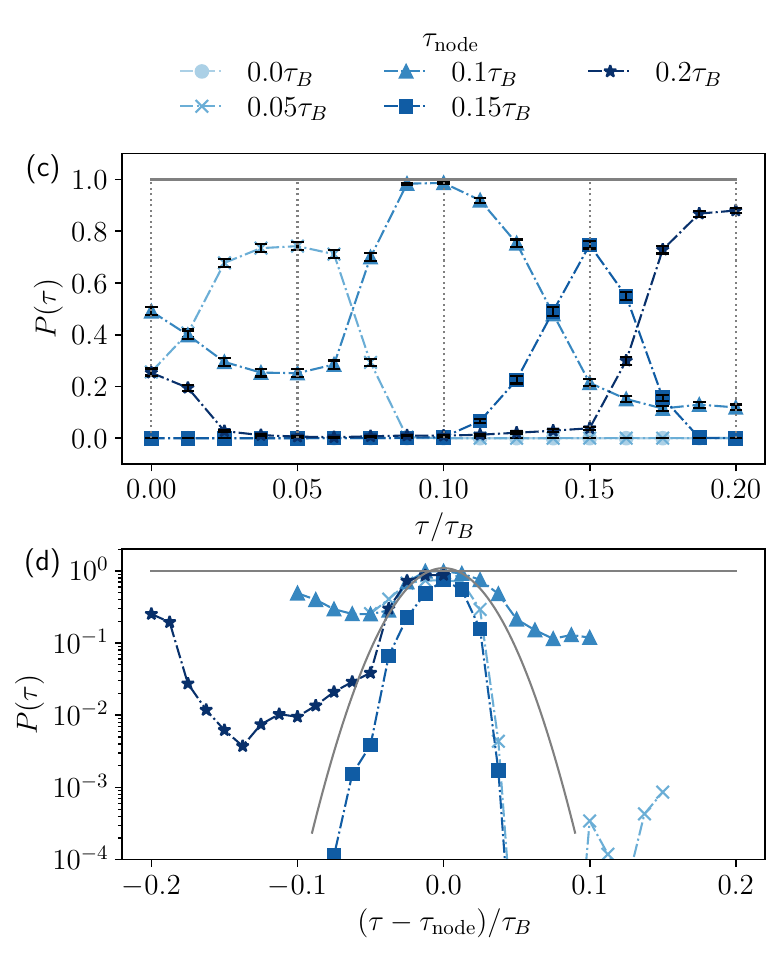}
    \caption{
    Inference results for a network trained on non-linear systems [Eq.~\eqref{eq:dw_linear}-\eqref{eq:tilted_washboard_nonlinear}] with white noise, for trajectories from colored noise systems based on [Eq.~\eqref{eq:dw_linear}-\eqref{eq:tilted_washboard_nonlinear}] at correlation times $t_\mathrm{corr}=0.001 \tau$ [panels a) and b)] and  $t_\mathrm{corr}=0.1 \tau$ [panels c) and d)]. 
    Panels (a) and (c): Symbols (color intensity increasing with $\tau_{\rm node}$): The probability of inferring a delay time at an exit node corresponding to $\tau_\mathrm{node}$. Average over inference results for ten different trajectories (prepared as discussed in Appendix~\ref{sec:cnn_data_prep}) for each parameter combination. Each data point is an average over 
    27 
    different 
    combinations 
    of the system parameters.
    Error bars indicate one standard error of the mean. Vertical dotted lines indicate $\tau_{\rm node}$. 
    Panels (b) and (d): Same data (but log-scaled $y$-axis) as in panel (a) and (c) represented by scaling and shifting the data to reflect the deviation from the output node $\tau_{\rm node}$. Grey Gaussian curve centered around zero as well as horizontal grey line (indicating probability one) act as guides for the eye. 
    }
    \label{fig:NN_colored_noise_inference}
\end{figure*}

For the neural network approach, Fig.~\ref{fig:NN_colored_noise_inference} shows data for inference results of the modified nonlinear systems 
in Eqs.~\eqref{eq:dw_linear}-\eqref{eq:tilted_washboard_nonlinear} at two different correlation times using a network trained solely on white noise systems. In the white noise limit, i.e., $t_\mathrm{corr}\ll\tau$, the inference results resemble those obtained for white noise systems. However, when the correlation time is of the order of the delay time or longer, the resulting trajectories look significantly different than before. 
Thus, as we expect, a neural network trained on systems with only white noise may not be able to accurately infer the time delay for systems with longer correlation times. Therefore, one requires to train the neural network with different correlation times (including different combinations of other parameters) to have better accuracy when inferring the time delay.  
\clearpage
%\bibliography{Inferringreferences_from_work}

\end{document}